\documentclass[10pt]{article}
\usepackage{epsf}
\usepackage{amsmath}
\usepackage{amssymb}
\usepackage{graphicx}

\newcommand{\sfrac}[2]{\mathchoice
  {\kern0em\raise.5ex\hbox{\the\scriptfont0 #1}\kern-.15em/
   \kern-.15em\lower.25ex\hbox{\the\scriptfont0 #2}}
  {\kern0em\raise.5ex\hbox{\the\scriptfont0 #1}\kern-.15em/
   \kern-.15em\lower.25ex\hbox{\the\scriptfont0 #2}}
  {\kern0em\raise.5ex\hbox{\the\scriptscriptfont0 #1}\kern-.2em/
   \kern-.15em\lower.25ex\hbox{\the\scriptscriptfont0 #2}}
  {#1\!/#2}}

\def\dt     {{\Delta t}}
\def\myhalf {\sfrac{1}{2}}
\def\nph    {{n+\myhalf}}
\def\Fb     {{\bf F}}
\def\Ub     {{\bf U}}
\def\Qb     {{\bf Q}}
\def\Sg     {{\bf S}_g}

\newcommand{\Dx}{\boldsymbol{\nabla}}
\newcommand{\vel}{\boldsymbol{u}}
\newcommand{\grav}{\boldsymbol{g}}
\newcommand{\ez}{\boldsymbol{\hat{e}}_z}
\newcommand{\Ttrip}{T_{\mathrm{trip}}}
\newcommand{\Eov}{e_{0v}}
\newcommand{\pvstar}{p_v^*}
\newcommand{\ptrip}{p_{\mathrm{trip}}}
\newcommand{\cva}{c_{va}}
\newcommand{\cvv}{c_{vv}}
\newcommand{\cvl}{c_{vl}}
\newcommand{\cvm}{c_{vm}}
\newcommand{\cpa}{c_{pa}}
\newcommand{\cpv}{c_{pv}}
\newcommand{\cpm}{c_{pm}}
\newcommand{\cpl}{c_{pl}}
\newcommand{\qvstar}{q_v^*}
\newcommand{\ehat}{\widehat{e}}
\newcommand{\aav}{\alpha_v}
\newcommand{\bbv}{\beta_v}
\newcommand{\Tguess}{\widetilde{T}}
\newcommand{\Told}{T_{\mathrm{old}}}
\newcommand{\tol}{tol}
\newcommand{\tsat}{t_{\mathrm{sat}}}
\newcommand{\dtsat}{\Delta t_{\mathrm{sat}}}
\newcommand{\rvs}{r_{vs}}
\newcommand{\Sov}{S_{0v}}
\def\ds{\displaystyle}

\title{A Numerical Study of Methods for Moist Atmospheric Flows: Compressible Equations}

\author{Max~Duarte\footnotemark[1] \and 
        Ann~S.~Almgren\footnotemark[1] \and 
        Kaushik~Balakrishnan\footnotemark[1] \and 
        John~B.~Bell\footnotemark[1]\and 
        David~M.~Romps\footnotemark[2]~\footnotemark[3]}

\begin{document}
\maketitle

\renewcommand{\thefootnote}{\fnsymbol{footnote}}
        
\footnotetext[1]{Center for Computational Sciences and Engineering,
                 Lawrence Berkeley National Laboratory,
                 Berkeley, CA 94720}
\footnotetext[2]{Dept. of Earth and Planetary Science,
                 University of California, Berkeley,
                 Berkeley, CA 94720}
\footnotetext[3]{Earth Sciences Division, 
                 Lawrence Berkeley National Laboratory,
                 Berkeley, CA 94720}
                 
\begin{abstract}
We investigate two common numerical techniques for integrating reversible
moist processes in atmospheric flows in the context of solving the
fully compressible Euler equations.  
The first is a one-step, coupled technique based
on using appropriate invariant variables 
such that terms resulting from phase change are eliminated in the governing equations.
In the second approach, which is a two-step scheme,
separate transport equations for liquid water and vapor water are used,
and no conversion between water vapor and liquid water is allowed in the
first step, while in the second step a saturation
adjustment procedure is performed that correctly allocates the
water into its two phases based on the Clausius-Clapeyron formula.
The numerical techniques we describe are first
validated by comparing to a well-established benchmark problem.
Particular attention is then paid to the effect of changing the time scale
at which the moist variables are adjusted to the saturation requirements
in two different variations of the two-step scheme.
This study is motivated by the fact that when acoustic modes are integrated separately 
in time (neglecting phase change related phenomena), or when sound-proof equations 
are integrated,  
the time scale for imposing saturation adjustment is typically much larger than the 
numerical one related to the acoustics.  
\end{abstract}

\section{Introduction}
A key issue in moist atmospheric flow modeling involves the 
interplay between the dynamics of the flow and the 
thermodynamics 
related to reversible and irreversible moist processes.
In this paper we focus on reversible processes, i.e. water phase changes, 
using an exact Clausius-Clapeyron formula for moist thermodynamics, 
and considering the effects of the specific heats of water
and the temperature dependency of the latent heat
(as in \cite{Satoh2003,Romps2008}).
Specifically, we want to characterize the impact of 
modifying the time scale at which the moist thermodynamics is
adjusted to the saturation requirements.

Atmospheric flow models are often cast in terms of the 
potential temperature and Exner function. For moist
atmospheres an equivalent potential temperature is
typically used as a prognostic variable; see, e.g., \cite{Klemp1978}.
Alternatively, \cite{Ooyama1990}, for example,
writes the equations of motion 
in terms of conserved variables,
while other thermodynamic variables such as pressure,
are recovered diagnostically.
This formulation was later extended to include
irreversible thermodynamic processes (like precipitation)
in \cite{Ooyama2001},
and inspired the development of some other conservative schemes
considering both primitive variables 
(e.g., \cite{Satoh2002,Satoh2003,Satoh2008})
and a potential temperature-type of formalism
(e.g., \cite{Klemp2007}).

We follow here the approach of \cite{Ooyama1990} in formulating 
the problem based on the separation of dynamics 
and thermodynamics. For the dynamics, we explicitly evolve the compressible Euler 
equations with time steps dictated by the acoustic CFL condition.
This allows us to focus on issues of how to couple the moist thermodynamic 
processes with the dynamics.
In particular we can modify the time step
associated with the moist thermodynamic adjustments without changing either
the formulation of or the numerical solution procedure for the dynamics.
In the same spirit, we write the Euler equations in conservation
law form (similar to \cite{Satoh2003}), which eases the inclusion of 
time-varying thermodynamic parameters for moist air.  

Within this formulation, we consider two numerical treatments of 
moist microphysics (see, e.g., \cite{Grabowski1990}).
In the first approach, often referred to as the invariant (conservative)
variables approach, the equations of motion are defined
using appropriate invariant variables such that terms 
resulting from phase change are eliminated in the governing equations,
while the remaining variables are diagnostically recovered
(cf. \cite{Hauf1987}).
(This is the case, for instance, in \cite{Ooyama1990,Ooyama2001}
for total water content and entropy.)
In the second approach, which is more common, the 
equations are again defined using the
conservative variable for moist energy,
but in this case separate transport equations for liquid water and 
water vapor are used.
The essence of the two-step scheme is that the dynamics are evolved
in the first step without allowing any conversion between
water vapor and liquid water. In the second step a saturation
adjustment procedure is performed that correctly allocates the
water into its two phases based on the Clausius-Clapeyron formula
(cf. \cite{Soong1973}).
Similar two-step schemes have been considered, for instance, in 
\cite{Klemp1978,Satoh2003}.

We explore two variants of the two-step scheme.  In the first variant,
even though liquid water and water vapor are advanced without 
accounting for phase change, a saturation adjustment procedure is used 
to diagnose thermodynamic variables such as pressure and the specific 
heat of moist air that are used to advance the dynamics during the first step.
In the second variant, the dynamics is evolved without any adjustment
of the water variables or the moist thermodynamics.

Two different questions concerning the time scale for saturation adjustment
can be addressed in this way.  Using the first variant,
we assess the impact of advancing liquid water and water vapor without 
phase change terms, but with moist dynamics equivalent to that
computed with the one-step coupled scheme.
Using the second variant, we investigate the impact of completely
separating the saturation adjustment from the dynamics.

\section{Governing equations}\label{sec:gov_eqs}
We begin by writing the fully compressible equations of motion
expressing conservation of mass, momentum, and energy 
in a constant gravitational field,
\begin{eqnarray}
\frac{\partial \rho}{\partial t} + \Dx \cdot \left( \rho \vel \right) 
&=& 0, \label{eqn:cont} \\
\frac{\partial \left( \rho \vel\right) }{\partial t} + 
\Dx \cdot \left( \rho \vel \vel \right ) + \Dx p & = & -\rho g\ez, 
\label{eqn:momentum_simple}\\
\frac{\partial \left( \rho E \right) }{\partial t} + 
\Dx \cdot \left( \rho  E \vel+ p\vel \right ) & = & 
- \rho g \left(\vel \cdot \ez \right),
\label{eqn:energy_simple}
\end{eqnarray}
in which we neglect Coriolis forces and viscous terms, 
as well as the influence of thermal conduction and radiation.
Here $\rho$ is the total density and $\vel$ is the velocity.
The energy, $E,$ is defined as the sum of internal plus kinetic energies, 
and the pressure, $p$, is defined by an equation of state (EOS).
We include gravitational acceleration given by
$\grav = -g \ez$, where $\ez$ is the unit vector in the vertical 
direction. 

We then follow the formalism as in \cite{Romps2008}
for moist atmospheres with the additional simplification
that at any grid point all phases have the same temperature and velocity.
Here we also ignore ice-phase microphysics,
precipitation fallout, and subgrid-scale turbulence.
We consider an atmosphere with three components, 
dry air, water vapor, and liquid water, and treat
moist air as an ideal mixture with the water phases in
thermodynamic equilibrium,
so that only reversible processes are taken into account.
Denoting by $q_a$, $q_v$, and $q_l$
the mass fraction of dry air, water vapor,
and liquid water, respectively, we write
\begin{eqnarray}
 \frac{\partial \left( \rho q_a \right)}{\partial t}  
+ \Dx \cdot \left( \rho q_a \vel \right) & = & 0, \label{eqn:qa}\\
\frac{\partial \left( \rho q_v \right)}{\partial t}  
+ \Dx \cdot \left( \rho q_v \vel \right) & =  & e_v, \label{eqn:qv}\\
\frac{\partial \left( \rho q_l \right)}{\partial t}  
+ \Dx \cdot \left( \rho q_l \vel \right) & = & -e_v,\label{eqn:ql}  \enskip .
\end{eqnarray}
Since $\rho$ is the total density (i.e. it includes dry air,
water vapor and liquid water), we have that $q_a+q_v+q_l = 1$. 
The evaporation rate, $e_v,$ has dimensions of mass per volume per time; 
negative values of $e_v$ correspond to condensation.
Introducing the mass fraction of total water,
$q_w =q_v +q_l$, equations (\ref{eqn:qv})--(\ref{eqn:ql})
can be also recast as
\begin{equation}\label{eqn:qw}
\frac{\partial \left( \rho q_w \right)}{\partial t}  
+ \Dx \cdot \left( \rho q_w \vel \right) = 0.
\end{equation}
The energy $E$ in (\ref{eqn:energy_simple}) is defined in this work as
\begin{equation*}\label{eqn:E_tot}
E = \ehat + \frac{\vel \cdot \vel}{2},
\end{equation*}
where 
$\ehat$ stands for the specific internal energy of
moist air.
The constant-volume specific heat of moist air is
given by
\begin{equation*}\label{eqn:cvmix}
\cvm = q_a \cva + q_v \cvv + q_l \cvl,
\end{equation*}
with constant specific heats at constant volume:
$\cva$, $\cvv$, and $\cvl$, for 
the three components: air, water vapor, and liquid water, respectively.
The internal energy of
moist air is thus defined as
\begin{equation}\label{eqn:ehat}
\ehat = \cvm \left(T - \Ttrip  \right) + q_v \Eov,
\end{equation}
where $\Ttrip$ is the triple-point temperature,
and $\Eov$ is the specific internal energy of water vapor
at the triple point.
(Following \cite{Romps2008}, we neglect the contribution of the
specific internal energy of dry air at the triple point 
in the definition of $\ehat$.)
\cite{Satoh2003} considered the same formulation
for the internal energy of moist air (\ref{eqn:ehat}), 
but included potential energy in the definition of total energy.
We note that the definition of $E,$ and in particular of $\ehat,$
yields an energy equation (\ref{eqn:energy_simple})
with no source terms related to phase changes.
In the case of a potential temperature-type of formalism,
this can be also achieved by defining
a liquid (or ice-liquid) potential temperature
as originally introduced by \cite{Betts1973,Tripoli1981},
and considered, for instance, in 
\cite{Tripoli1992,Walko2000,Jiang2000,Walko2008}.
 
An equation of state for moist air must be provided to close the system.
For the sake of illustration, we consider in this study
a standard approach adopted in atmospheric flows in which
dry air and water vapor are treated as ideal gases
(see, e.g., \cite{Ooyama1990,Satoh2003,Klemp2007}).
The partial pressures of dry air and water vapor are then given by
$p_a = \rho q_a R_a T$ and 
$p_v = \rho q_v R_v T,$
where 
$R_a$ and $R_v$ are the specific gas constants
for dry air and water vapor, respectively.
Denoting by $M_a$ and $M_v$ the molar masses of dry air and water,
respectively, we know that
$R_a = R/M_a$ and $R_v = R/M_v$, where 
$R$ is the universal gas constant for ideal gases.
If we define the specific gas constant of moist air as
\begin{equation*}\label{eqn:R_m}
R_m = q_aR_a + q_vR_v = \left(\frac{q_a}{M_a} + \frac{q_v}{M_v}\right) R,
\end{equation*}
then the sum of the partial pressures
defines the total pressure of a parcel,
\begin{equation}\label{eqn:pres}
p = p_a + p_v = \rho R_m T \enskip .
\end{equation}
Additionally, the specific heat capacities at constant
pressure can be defined as
\begin{equation*}\label{eqn:cpm}
\cpa = \cva + R_a, \quad
\cpv = \cvv + R_v, \quad
\cpm = \cvm + R_m,
\end{equation*}
for dry air, water vapor, and moist air, respectively.
A common approximation in cloud models 
is to neglect the specific heats of water vapor and liquid water
(see, e.g., \cite{Bryan2002} for a study and discussion on this topic).
Here we consider specific heats for all three phases.

Now, the saturation vapor pressure 
with respect to liquid water, $\pvstar$, is defined 
by the following Clausius-Clapeyron relation:
\begin{equation}\label{eqn:pvstar}
\pvstar(T) = \ptrip \left(\frac{T}{\Ttrip}\right)^{\aav}
\exp \left [ \bbv 
\left( \frac{1}{\Ttrip}-\frac{1}{T} \right) \right ],
\end{equation}
with constants $\aav$ and $\bbv$, given, for instance, by 
\begin{equation}\label{eqn:constant_pvstar}
\aav = \frac{\cpv-\cvl}{R_v}, \qquad 
\bbv = \frac{\Eov -(\cvv-\cvl)\Ttrip}{R_v},
\end{equation}
as in \cite{Romps2008}.
The saturated mass fraction of water vapor,
$\qvstar$, can be then computed from the EOS,
given in this case by
\begin{equation}\label{eqn:qvstar}
\qvstar(\rho,T) = \frac{\pvstar}{\rho R_v T} \enskip .
\end{equation}
Following \cite{Ooyama1990,Satoh2003}, we assume 
that air parcels cannot be supersaturated,
and thus water vapor mass fraction, $q_v$, 
cannot exceed its saturated value, $\qvstar$.

\section{Numerical Methodology}\label{Sec:Num_Met}
In what follows we describe the numerical methodology we use
to solve equations (\ref{eqn:cont})--(\ref{eqn:ql})
for moist flows.  The detailed numerics for dry flow are as described 
in \cite{CASTROpaper}, which describes the CASTRO code,
a multicomponent compressible flow solver.
Our attention here will be mainly focused on the 
incorporation of moist reversible processes, and we discuss how
the different approaches handle phase transitions within the 
numerical solution of the overall flow dynamics.
We will refer to the first as the one-step coupled scheme, in which 
the solution variables include energy of moist air and total water,
and the effects of phase change are diagnostically evaluated and
incorporated when computing the dynamics within each time step.
Two variants of the two-step technique will be then studied.
In the 
first one, denoted the two-step semi-split scheme,
the density, momentum and energy are evolved exactly as in the 
one-step fully coupled scheme, but liquid water and water vapor
are advected separately and no conversion between them is allowed.
In the second one, 
denoted the two-step fully-split scheme,
the dynamics are first evolved neglecting any effects of phase change.
In both of the split schemes, the first step is used to
advance the solution by one or more time steps before being
followed by an adjustment procedure that imposes the
saturation requirements using the Clausius-Clapeyron formula, 
specifically updating $q_l$, $q_v$, and $T$.
Notice that during the first step 
the water phases may not be in thermodynamic equilibrium
anymore; therefore, the
saturation adjustment naturally involves an irreversible process.
The latter is, however, a result of the numerical approach
to approximate moist flows with phase transitions,
which are considered as reversible processes in our model.
Our formulation and implementation of moist microphysics for the 
first and 
second two-step schemes
are similar, respectively, to 
\cite{Ooyama1990} and \cite{Satoh2003}.

In all three cases we define a state vector of conserved variables,
$\Ub,$ and write 
the time evolution of $\Ub$ in the form
\begin{equation*}
\frac{\partial\Ub}{\partial t} = -\nabla\cdot\Fb + \Sg,
\end{equation*}
using a finite volume discretization,
where $\Fb$ is the flux vector and $\Sg$ represents only the
gravitational source terms in the equations for momentum and
energy.  We advance $\Ub$ by one time step, $\Delta t,$
using the time discretization,
\begin{equation}\label{eqn:time_disc}
\Ub^{n+1} = \Ub^{n} - \Delta t \nabla \cdot\Fb^\nph + \Delta t \Sg^\nph.
\end{equation}
The total density, $\rho$, as well as $\rho q_a$ and $\rho q_w$
(or $\rho q_a, \rho q_v$ and $\rho q_l$),  are included in $\Ub$;
following the advective update we adjust $q_a$ and $q_w$
to enforce that $\rho = \rho q_a + \rho q_w$ 
(or equivalently $\rho = \rho q_a + \rho q_v + \rho q_l$).
The construction of $\Fb$ is purely explicit, and based on an unsplit
Godunov method with characteristic tracing.  
The solution, $\Ub,$ is defined on cell centers;
we predict primitive variables, 
$\Qb,$
from cell centers at time $t^n$ to edges at time $t^{\nph}$, and use an
approximate Riemann solver to construct fluxes $\Fb^\nph$ on cell faces.
Within the construction of the fluxes, the pressure is diagnostically
computed as needed on cell edges using the EOS in (\ref{eqn:pres}).
As we will see below, the schemes differ in the values of the
moist thermodynamic variables that enter this intermediate call to the EOS.
This algorithm is formally second-order in both space and time;
we refer to \cite{CASTROpaper} for the complete details of 
this numerical implementation.

The time step in (\ref{eqn:time_disc})
is computed using the standard CFL condition for explicit methods.  
Following \cite{CASTROpaper},
we set a CFL factor $\sigma^\mathrm{CFL}$
between 0 and 1,
and for a calculation in $n_\mathrm{dim}$ dimensions, 
\begin{equation}\label{eqn:dtCFL}
\dt = \sigma^\mathrm{CFL}  \min_{i=1\ldots n_\mathrm{dim}} \left \{ \dt_i \right \},
\qquad
\dt_i = \frac{\Delta x_i}{|{\vel}_i| + c_m \; },
\end{equation}
with $c_m$, the sound speed in moist air, and $\dt_i$
computed as the minimum over all cells.
The sound speed is computed using the moist EOS,
and is defined in this study as for an ideal gas:
\begin{equation*}\label{eqn:cs}
c_m = \sqrt{\frac{\gamma_m p}{\rho}}, \qquad
\gamma_m = \frac{\cpm}{\cvm},
\end{equation*}
where $\gamma_m$ is the isentropic expansion factor of moist air.

In each of the schemes we also need to be able to obtain 
point-wise values of $(q_v,q_l,T)$ given $(\rho,\vel,E,q_a,q_w)$,
using the Clausius-Clapeyron relation and the saturation requirements.
We refer to this as the saturation adjustment procedure, and 
do so by solving the following nonlinear system of equations \cite{Satoh2003}:
\begin{equation}\label{eqn:solveT}
\left.
\begin{array}{l}
\ds \ehat = E - \frac{\vel \cdot \vel}{2} =
\cvm(q_a,q_v,q_l)  \left(T - \Ttrip  \right) + q_v \Eov, \\[1.5ex]
q_v = \min\left[\qvstar(\rho,T),q_w\right],  \\[1.5ex]
q_l = q_w - q_v.
\end{array}
\right\}
\end{equation}

The numerical solution of (\ref{eqn:solveT}) 
uses an iterative Newton solver, 
described in detail here for the sake of completeness:
\begin{description}
\item[Step 1:] {\em Initialization.}
Define the initial guess, $\Tguess = \Told$, 
where $\Told$ is the last known temperature in the current cell.
\item [Step 2:] {\em Compute mass fractions: $q_v$ and $q_l$.}
Following the Clausius-Clapeyron relation (\ref{eqn:pvstar}), compute
\begin{equation*}\label{eqn:pvstartilde}
\widetilde{\pvstar}\left(\Tguess\right) = \ptrip \left(\frac{\Tguess}{\Ttrip}\right)^{\aav}
\exp \left [ \bbv 
\left( \frac{1}{\Ttrip}-\frac{1}{\Tguess} \right ) \right ],
\end{equation*}
and $\widetilde{\qvstar} = \widetilde{\qvstar}(\rho,\Tguess)$
from (\ref{eqn:qvstar}),
so that
\begin{equation*}
\widetilde{q_v} = \min\left[\widetilde{\qvstar},q_w\right], \qquad
\widetilde{q_l} = q_w -\widetilde{q_v} \enskip .
\end{equation*}
We can then evaluate
\begin{equation*}\label{eqn:ehattilde}
\widetilde{\ehat} =
\cvm(q_a,\widetilde{q_v},\widetilde{q_l})  
\left(\Tguess - \Ttrip  \right) + \widetilde{q_v} \Eov.
\end{equation*}
\item[Step 3:]  {\em Update temperature: $T$.}
Define a local function: $f(\Tguess) = \widetilde{\ehat} - \ehat$,
and update $\Tguess$ by 
computing a Newton correction step:
\begin{equation*}\label{eqn:newton_T}
\Tguess = \Tguess - \Delta \Tguess, \qquad
\Delta \Tguess = 
f (\Tguess )/\partial_{\Tguess} f(\Tguess),
\end{equation*}
where
\begin{equation*}\label{eqn:derfT}
\partial_{\Tguess} f =
\partial_{\Tguess} \qvstar
(L_e(\Tguess) - R_v\Tguess) + \cvm,
\
\partial_{\Tguess} \qvstar = 
\qvstar \left( \frac{\aav -1}{\Tguess} + \frac{\bbv}{ {\Tguess}^2 } \right),
\end{equation*}
with the latent heat of vaporization, $L_e$,
defined as
\begin{equation}\label{eqn:Le}
L_e(T) = \Eov + R_v T + (\cvv - \cvl)(T - \Ttrip).
\end{equation}
\item[Step 4:] {\em Stopping criterion.}
Introducing an accuracy tolerance, $\tol$, 
and denoting $err = |\Delta \Tguess/\Tguess |$,
we define the following stopping criterion:
\begin{itemize}
 \item If $err > \tol$: go back to {\bf Step 2};
 \item If $err \leq \tol$: stop iterating and set 
$T=\Tguess$, $q_v= \widetilde{q_v}$, and $q_l= \widetilde{q_l}$.
\end{itemize}

\end{description}
Notice that
if $q_w < \qvstar $, all water is in the form of vapor,
that is, $q_v = q_w$ and $q_l = 0$;
the temperature is hence directly computed from (\ref{eqn:ehat}),
or equivalently from {\bf Steps 1-4} considering that
in this case: $\partial_{\Tguess} f = \cvm$.
This procedure remains valid for any moist equation of state, as
long as a Clausius-Clapeyron relation (\ref{eqn:pvstar}) is available
to define the saturation pressure.

We note that for flows in which no phase change occurs, the 
time evolution of the solution in the one-step and two-step schemes 
will be identical.

\subsection{One-step Coupled Scheme}\label{Subsec:Coupled}
We consider the following set of evolution equations:
\begin{equation}\label{eqn:system1}
\left.
\begin{array}{rcl}
\ds \frac{\partial \rho}{\partial t} + \Dx \cdot \left( \rho \vel \right) &=& 0,\\[1.5ex]
\ds \frac{\partial \left( \rho \vel\right) }{\partial t} + 
\Dx \cdot \left( \rho \vel \vel \right ) + \Dx p &=& -\rho g\ez,\\[1.5ex]
\ds \frac{\partial \left( \rho E \right) }{\partial t} + 
\Dx \cdot \left( \rho E \vel  + p\vel \right ) &=& 
- \rho g \left(\vel \cdot \ez \right),\\[1.5ex]
\ds \frac{\partial \left( \rho q_a \right)}{\partial t}  
+ \Dx \cdot \left( \rho q_a \vel \right) &=& 0,\\[1.5ex]
\ds \frac{\partial \left( \rho q_w \right)}{\partial t}  
+ \Dx \cdot \left( \rho q_w \vel \right) &=& 0,
\end{array}
\right\}
\end{equation}
and close the system with the moist EOS (\ref{eqn:pres}).
\cite{Ooyama1990} considers the same formulation but
the conservation equation for entropy density of moist air 
is considered instead of $(\rho E)$.

We define the state vector of conserved variables, 
$\Ub = (\rho,\rho\vel,\rho E,\rho q_a,\rho q_w);$ 
the primitive variables in the flux construction are then
$\Qb = (\rho, \vel, \rho \ehat, q_a, q_w).$ 
In defining the pressure used to construct the fluxes 
we solve (\ref{eqn:solveT}) for $T,$ $q_v$, and $q_l$,
given the 
values of $\Qb$ before calling the EOS.
This approach is coupled in the sense that the moist processes are 
incorporated as part of the dynamical evolution of the system.  
Because we evolve $q_w$, rather than $q_v$ and $q_l$ separately,
and call the saturation adjustment procedure any time $q_v$ and $q_l$
are needed, there is never any lagging or neglect of moist effects.

\subsection{Two-step Schemes}\label{Subsec:Split}

Here we consider the following set of equations:
\begin{equation}\label{eqn:system2}
\left.
\begin{array}{rcl}
\ds \frac{\partial \rho}{\partial t} + \Dx \cdot \left( \rho \vel \right) 
& = &0,\\[1.5ex]
\ds \frac{\partial \left( \rho \vel\right) }{\partial t} + 
\Dx \cdot \left( \rho \vel \vel \right ) + \Dx p & =& -\rho g\ez,\\[1.5ex]
\ds \frac{\partial \left( \rho E \right) }{\partial t} + 
\Dx \cdot \left( \rho E \vel + p\vel \right ) &=& - \rho g \left(\vel \cdot \ez \right),\\[1.5ex]
\ds \frac{\partial \left( \rho q_a \right)}{\partial t}  
+ \Dx \cdot \left( \rho q_a \vel \right)& = &0,\\[1.5ex]
\ds \frac{\partial \left( \rho q_v \right)}{\partial t}  
+ \Dx \cdot \left( \rho q_v \vel \right) &= &e_v,\\[1.5ex]
\ds \frac{\partial \left( \rho q_l \right)}{\partial t}  
+ \Dx \cdot \left( \rho q_l \vel \right) &= &-e_v,
\end{array}
\right\}
\end{equation}
where we now define
$\Ub = (\rho,\rho\vel,\rho E,\rho q_a,\rho q_v, \rho q_l)$ 
and $\Qb = (\rho, \vel, \rho \ehat, q_a, q_v, q_l).$ 
We note that, in contrast to the one-step coupled scheme,
here we separately advance water vapor and liquid water
rather than advancing total water.
\cite{Satoh2003} considers the same setup
but an equation for internal energy accounting
only for sensible heat is evolved instead of $(\rho E)$;
in that formulation a source term corresponding to the 
latent heat release then appears in the conservation equation for energy.
In our case only the equations for $q_v$ and $q_l$
explicitly contain information about the water phase transitions,
which simplifies the comparison of the one- and two-step 
schemes for the purposes of the present study.

In the first step of both split schemes, $q_v$ and $q_l$ are
advected with $e_v = 0.$ In the semi-split scheme, the saturation adjustment process is performed before
the intermediate pressure is computed from the EOS to define $\Fb^{\nph}$, just as in 
the one-step coupled scheme; the only difference between
the procedure here and in the one-step scheme is that we must first define
$q_w = q_v + q_l$ before doing the saturation adjustment.  
In the fully-split scheme, the temperature and pressure are 
computed for $\Fb^{\nph}$ given the existing values of $q_v$ and $q_l$ on the faces; no 
saturation adjustment is performed.  
Given $\Fb^{\nph},$ the update in (\ref{eqn:time_disc}) is performed exactly
as in the one-step scheme.
In the second step of the split schemes we impose the saturation adjustment to 
correct $\Ub,$ specifically $q_v$ and $q_l$, but only if the designated time
interval has passed.

In both split schemes, the first step may be performed multiple times
before the second step is called. Defining $\tsat$ and $\dtsat$, respectively, 
as the time at which the saturation adjustment step is performed, 
and the specified time interval between saturation adjustments,
we can describe each of these schemes below.

\subsubsection{Two-step Semi-Split Scheme}\label{Subsubsec:SemiSplit}

\begin{description}
\item [Step I:] {\em Advance dynamics through $\dt$.} 
Advance (\ref{eqn:system2}) in time from $t^n$ to $t^{n+1} = t^n+\dt,$
advancing $q_v$ and $q_l$ with $e_v = 0,$  but define 
$q_w = q_v + q_l$ at $t^\nph$ to be used in the saturation adjustment procedure,  
and compute the intermediate pressure used to construct the fluxes with 
saturation-adjusted variables.

\item [Step II:] {\em Moist microphysics adjustment.}
If $t^{n+1} \geq \tsat + \dtsat$, 
where $\tsat$ records the last time the correction step was computed,
solve (\ref{eqn:solveT}) 
for $q_v$, and $q_l$ at $t^{n+1}$ given values of 
$(\rho,\ehat,q_a,q_w=q_v+q_l)$ at $t^{n+1},$
using the iterative procedure described in {\bf Steps 1-4}.
Set $\tsat = t^{n+1}$.
\end{description}

\subsubsection{Two-step Fully-Split Scheme}\label{Subsubsec:FullySplit}

\begin{description}
\item [Step I:] {\em Advance dynamics through $\dt$.}
Advance (\ref{eqn:system2}) in time from $t^n$ to $t^{n+1} = t^n+\dt$
with $e_v = 0.$
In defining the pressure used to construct the fluxes,
do not perform the saturation adjustment procedure.
Instead, since we explicitly evolve $q_v$ and $q_l$ separately,
compute the temperature, $T,$ directly from (\ref{eqn:ehat}) (given $E$ and $\vel$,
hence $\ehat$),
effectively neglecting any phase change that might occur during the time step.
The pressure is then determined from the EOS given these values.

\item [Step II] is exactly as above.
\end{description}

Notice that if $\dtsat \leq \dt$, the moist and thermodynamic variables
are corrected immediately after each update of the dynamics.
In the semi-split scheme, the dynamics are evolved with saturation-adjusted
variables, but $q_v$ and $q_l$ themselves drift from their correct values
at the end of the time step;
the larger $\dtsat$ is, the more they differ from those 
diagnostically recovered from the one-step solution.
In the fully-split scheme, the larger $\dtsat$ is,
the more the dynamics evolve neglecting phase changes.
Recall that our numerical implementation does not 
discriminate between fast and slow modes associated with the
compressible equations; therefore, whenever $\dtsat > \dt$,
where $\dt$ is limited by the acoustic CFL condition,
several dynamical time steps $\dt$ are performed 
before the saturation adjustment.

\section{Numerical Simulations}\label{Subsec:NumSim}

In what follows we first consider the benchmark problem
proposed in \cite{Bryan2002} for moist flows,
along with the corresponding configuration for dry air
originally presented in \cite{Wicker1998}.
Both cases are presented 
and results are compared with those obtained in \cite{Bryan2002}
in order to first validate our basic numerical implementation.
We then compare the approximations obtained with the different numerical 
schemes previously described.  In particular, we investigate the impact
of the time interval of saturation adjustment, $\dtsat,$ on the moist flow
for the two split schemes.
A second configuration based on \cite{GrabowskiClark1991}
is also studied for non-isentropic background states
and both saturated and only partially saturated media,
to further assess the different numerical techniques.

\subsection{Numerical Validation}\label{Subsec:NumVal}

\cite{Bryan2002} present solutions of a benchmark test case 
using the fully compressible equations,
where the conservation equations for 
water vapor and liquid water are written in terms of
the water vapor and cloud mixing ratios: $r_v = q_v/q_a$ and
$r_c = q_l/q_a$, respectively.  The conservation equation for 
energy (\ref{eqn:energy_simple})
is replaced by \begin{equation*}\label{eqn:DTDt}
\rho \cvm \left(
\frac{\partial T}{\partial t} +
\vel \cdot \Dx T
\right)
=
- p \left(\Dx \cdot \vel\right) - ( L_v - R_v T )e_v,
\end{equation*}
with the latent heat of vaporization $L_v$
defined as
\begin{equation}\label{eqn:Lv}
L_v = L_{v0} - (\cpl - \cpv)(T - T_0),
\end{equation} 
where $L_{v0}$ and $T_0$ are constant reference
values of $L_v$ and $T$, respectively.
The nondimensional Exner pressure, $\pi,$ and
potential temperature, $\theta,$ are used in \cite{Bryan2002},
defined as 
\begin{equation}\label{eqn:pi_theta}
\pi = \left( 
\frac{p}{p_{00}}
\right)^{R_a/\cpa},
\qquad
\theta = \frac{T}{\pi},
\end{equation}
where $p_{00}=1000\,$mb.
The numerical scheme thus solves time-dependent equations for
$(\vel,\pi,\theta,r_v,r_c)$, where the evaporation rate $e_v$
appears in the source terms for the equations for 
$\pi$, $\theta$, $r_v$, and $r_c$.
The technique introduced in 
\cite{Klemp1978} is used to integrate the equations
in two steps:
a dynamical step and the microphysics step.
In the dynamical step, $e_v$ is neglected and the 
portions of the governing equations that support acoustic waves 
are updated with a smaller time step than the other terms.
The model is integrated with a third-order Runge-Kutta 
scheme and fifth-order spatial discretization for the
advective terms.
Then, a saturation adjustment technique, similar to 
that proposed by \cite{Soong1973},
is used in the microphysics step in which only the terms 
involving phase change are included.
Notice that this approach is similar to
our fully-split
procedure described in \S~\ref{Subsubsec:FullySplit},
with the main difference that in our formulation 
the terms related to phase changes appear only in the equations for
$q_v$ and $q_l$.

The hydrostatic base state pressure can be found through
\begin{equation}\label{eqn:pi0}
\frac{d \pi_0}{d z} = - \frac{g}{\cpa\theta_{\rho 0}},
\end{equation}
where the subscript ``0'' stands for hydrostatic base quantities,
and the density potential temperature, $\theta_\rho,$
is defined as
\begin{equation}\label{eqn:thetarho}
\theta_\rho = \theta \frac{(1+r_v/\epsilon)}{(1+r_t)},
\end{equation}
with the total water mixing ratio, $r_t = q_w/q_a$,
and $\epsilon = R_a/R_v = M_v/M_a$.

For the next set of computations
we consider the following constant parameters, 
taken from \cite{Bryan2002}:
$R_a = 287\,$J kg$^{-1}$ K$^{-1}$,
$R_v = 461\,$J kg$^{-1}$ K$^{-1}$,
$L_{v0} = 2.5\times 10^{6}\,$J kg$^{-1}$,
$\cva = 717\,$J kg$^{-1}$ K$^{-1}$,
$\cvv = 1424\,$J kg$^{-1}$ K$^{-1}$,
$\cpl = 4186\,$J kg$^{-1}$ K$^{-1}$,
$T_0 = 273.15\,$K, and 
$g = 9.81\,$m s$^{-1}$.
The remaining parameters used in our model
are defined such that we
have the same definition 
of the latent heat of vaporization, that is,
$L_v=L_e$ from (\ref{eqn:Lv}) and 
(\ref{eqn:Le}).
Therefore
we just need to consider:
$\Ttrip=T_0$, $\cvl=\cpl$, and
$\Eov = L_{v0} - R_v\Ttrip$.
The saturation vapor pressure is computed 
with the Clausius--Clapeyron equation (\ref{eqn:pvstar})
with constants:
$\aav = 0$ and $\bbv = L_{v0}/R_v$,
with $\ptrip = 611\,$Pa,
taken from \cite{ONeill2013} that considers
the same benchmark problem.

\subsubsection{The Dry Simulation}

Following \cite{Wicker1998} and \cite{Bryan2002},
we consider a two-dimensional computational 
domain with height $10\,$km and width $20\,$km.
The initial atmospheric
environment is defined by a constant potential
temperature of $\theta_0=300\,$K,
and the pressure field is obtained by integrating
upwards the hydrostatic equation (\ref{eqn:pi0}).
A warm perturbation is introduced in the domain, 
given by 
\begin{equation}\label{eqn:dry_pert}
\theta' = 2 \cos ^2 \left( \frac{\pi L}{2} \right),
\end{equation}
where
\begin{equation}\label{eq:pert_L}
L = \min \left\{
1,\sqrt{\left(\frac{x-x_c}{x_r} \right)^2
+
\left(\frac{z-z_c}{z_r} \right)^2
}
\right\},
\end{equation}
with $x_c = 10\,$km,
$z_c = 2\,$km, and
$x_r = z_r = 2\,$km.
Notice that our formulation does not use
the $\theta - \pi$ formalism; the expressions 
(\ref{eqn:pi_theta}) are used for the conversions,
while the initial perturbation (\ref{eqn:dry_pert})
is applied at constant pressure $p(z)$.
We impose zero normal velocities and homogeneous
Neumann boundary conditions for the tangential velocity
components on all four boundaries. 
(Tangential velocity boundary conditions are necessary 
for the unsplit computation of advective fluxes
in this specific numerical solver \cite{CASTROpaper}.)
For the thermodynamic variables, we impose
homogeneous Neumann boundary conditions on the horizontal sides;
the background state is reconstructed by extrapolation at 
vertical boundaries in order to determine the corresponding fluxes.
\begin{figure}[!ht]
\noindent
\includegraphics[width=0.49\textwidth]{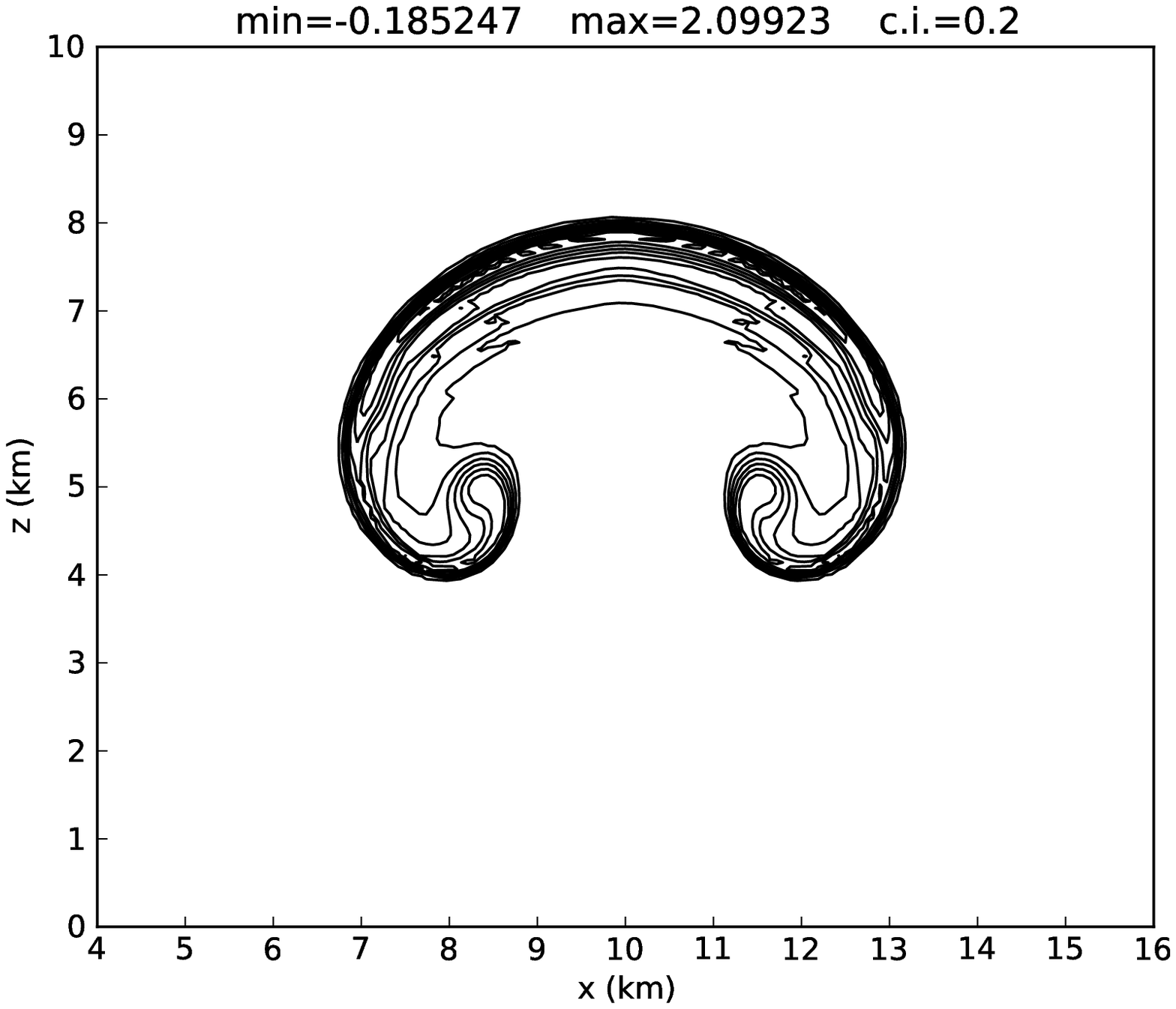}
\includegraphics[width=0.49\textwidth]{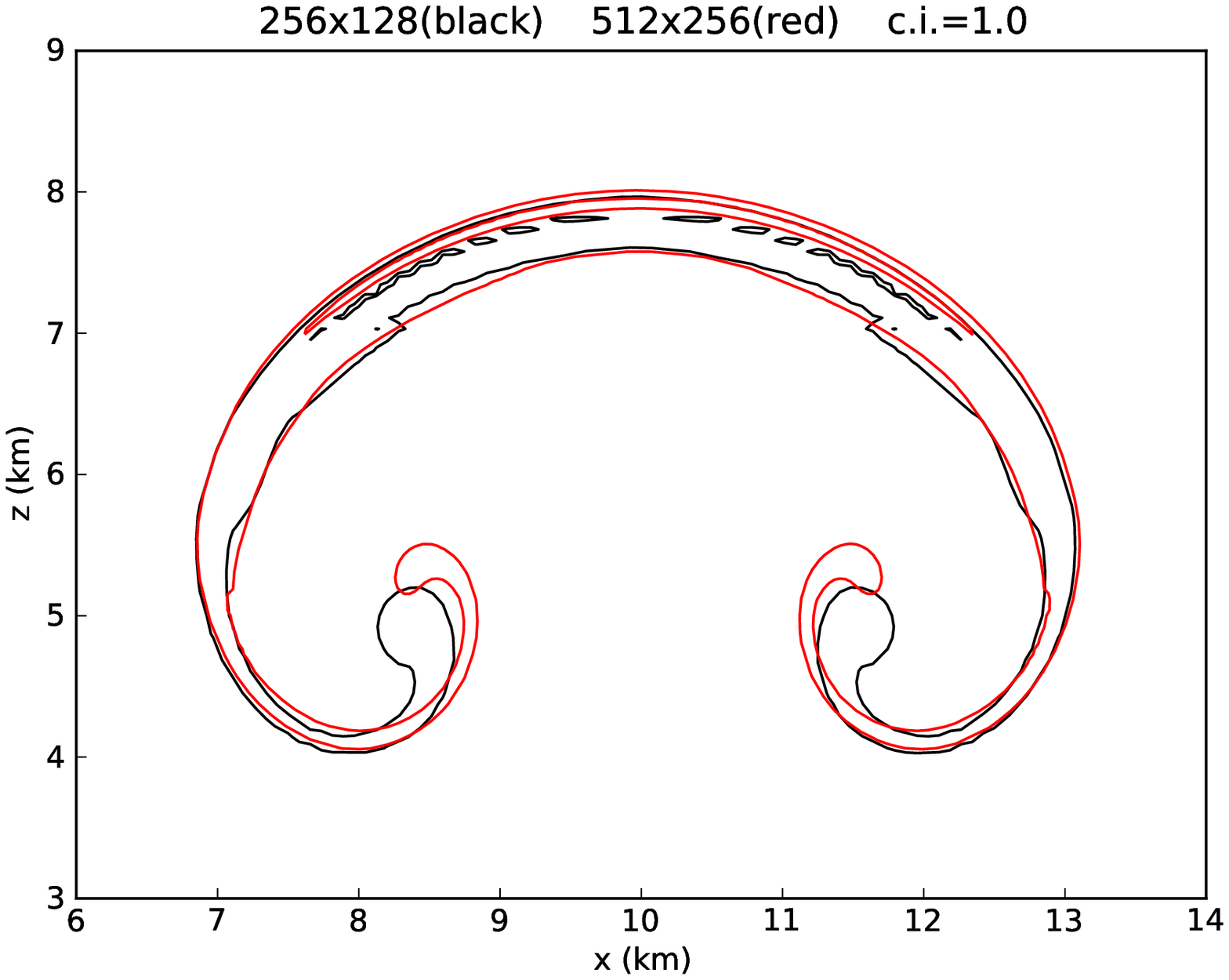}\\
\caption{Dry thermal simulation at $1000\,$s.
Left: perturbation potential temperature on a $256\times 128$ grid,
contoured every $0.2\,$K.
Right: comparison of the perturbation potential temperature
computed on a $256\times 128$ (black) and $512\times 256$ (red) grids,
contours every $1\,$K.}
\label{fig:dry_128}
\end{figure}
\begin{figure}[!ht]
\noindent
\includegraphics[width=0.49\textwidth]{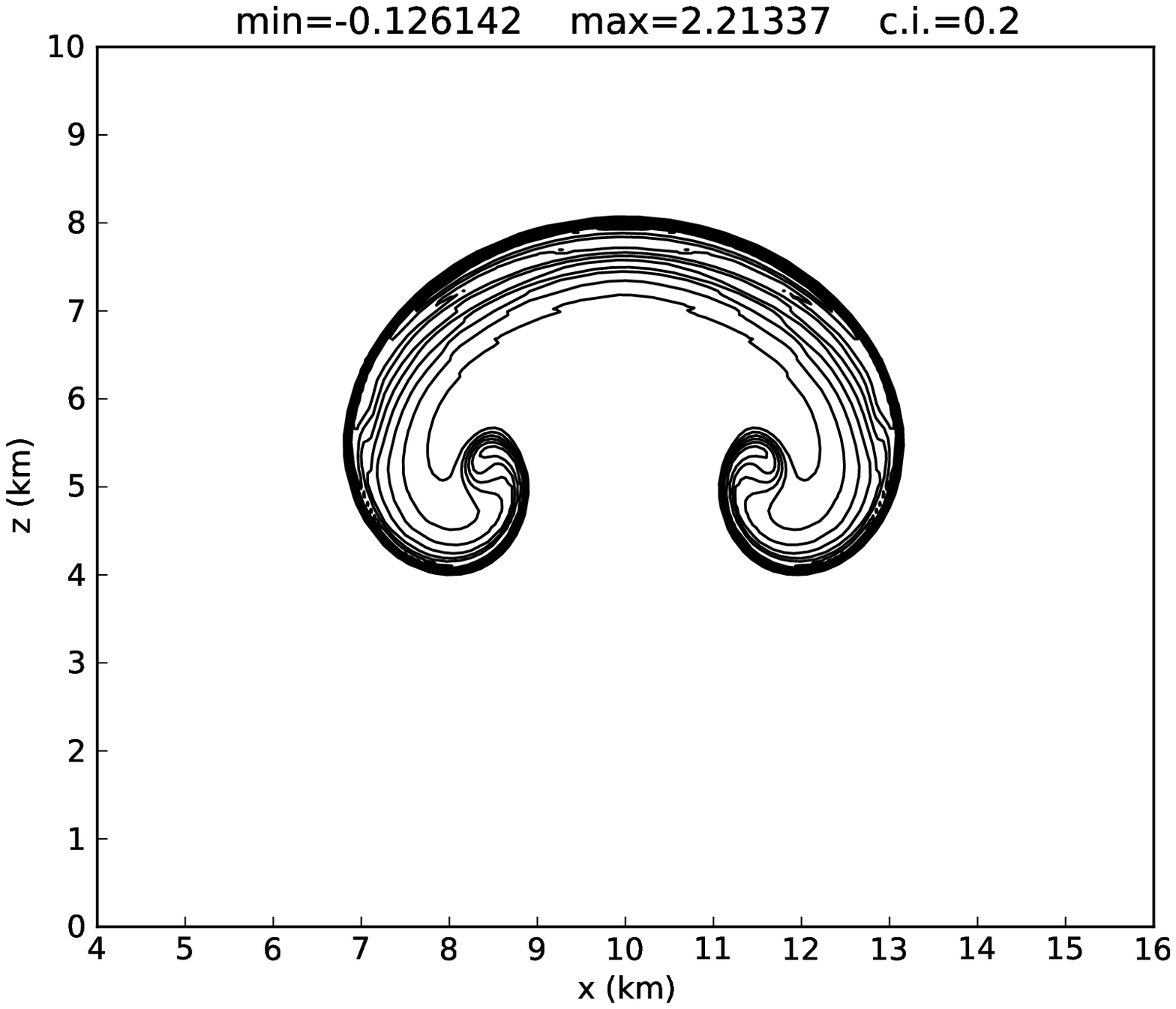}
\includegraphics[width=0.49\textwidth]{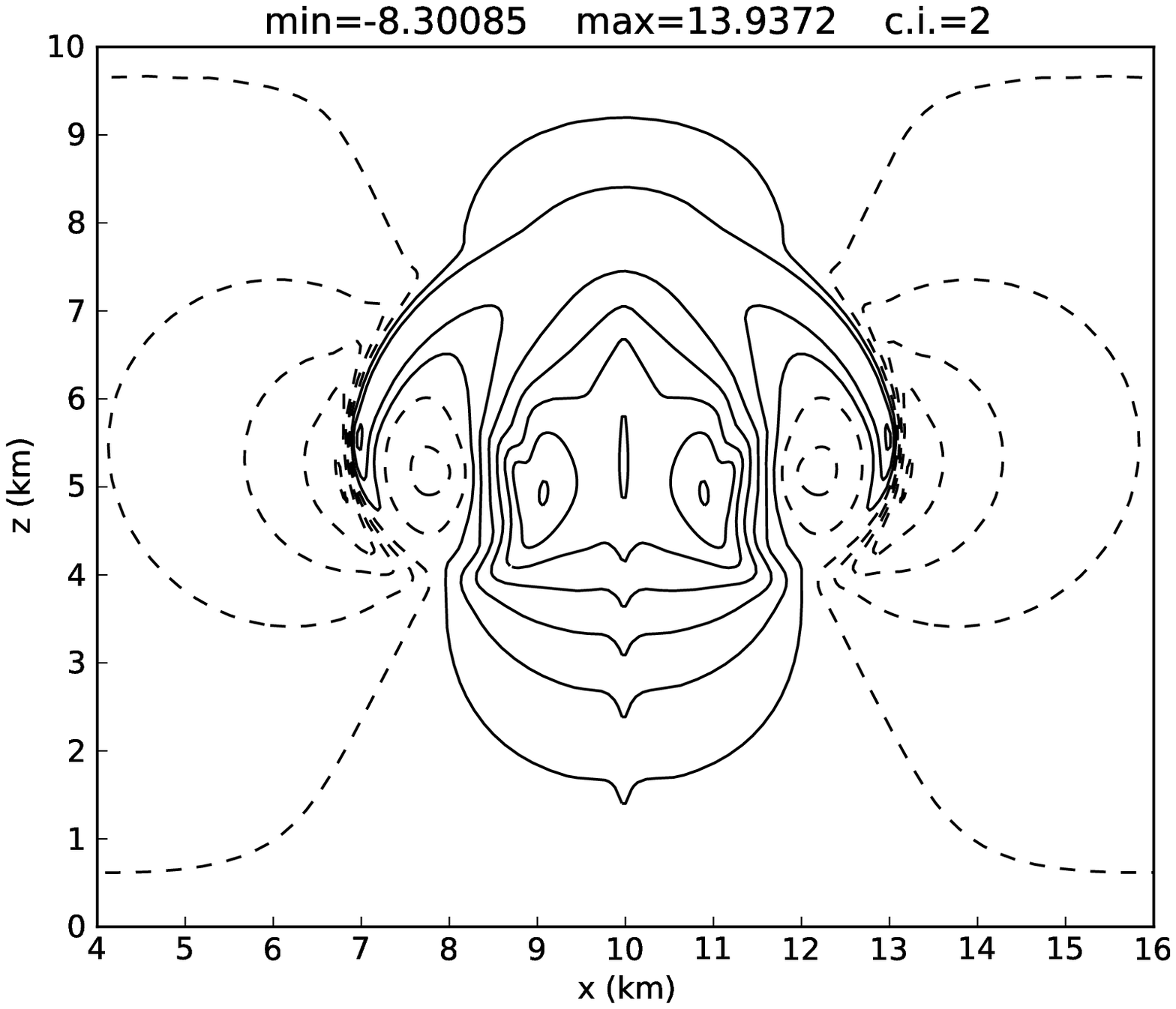}\\
\caption{Dry thermal simulation at $1000\,$s on a $512\times 256$ grid.
Left: perturbation potential temperature
contoured every $0.2\,$K.
Right: vertical velocity contoured every $2\,$m s$^{-1}$,
negative contours are dashed.
Contrast to Fig.~1 in \cite{Bryan2002}.}
\label{fig:dry_256}
\end{figure}

Let us first consider
a uniform grid of $256\times 128$ points,
slightly finer than the original $100\,$m
grid spacing in \cite{Bryan2002}.
For all computations the time step is computed using
the CFL factor $\sigma^\mathrm{CFL}=0.9$ in (\ref{eqn:dtCFL}).
This yields roughly constant time steps of about $0.2\,$s
in this configuration for the dry thermal computations.
Figure \ref{fig:dry_128} (left)
illustrates the numerical results 
for the perturbation potential temperature
($\theta'=\theta - \theta_0$) 
after $1000\,$s.
The maximum and
minimum values for $\theta'$ are 
given by $2.09923\,$K and $-0.18525\,$K,
respectively, compared with the original 
$2.07178\,$K and $-0.144409\,$K in \cite{Bryan2002}.
Good general agreement is also found with respect 
to the solutions in \cite{Bryan2002}
in terms of the height and width of the rising thermal.
However, some differences in the dynamics can be noticed
around the two vortices developed on the
sides of the thermal.
In \cite{Bryan2002} both tips of the thermal
seem to roll up slightly higher around the vortex cores.
The maximum and minimum values of vertical velocity are in fact
localized in this region, which in our computation are given by
$12.4991\,$m s$^{-1}$ and $-7.58896\,$m s$^{-1}$,
respectively, slightly lower than the values of 
$14.5396\,$m s$^{-1}$ and $-8.58069\,$m s$^{-1}$
in \cite{Bryan2002}.

Besides the different choice of variables,
there are two main differences between our implementation
and the one in \cite{Bryan2002} that may
explain the height difference of the thermal tips.
The first difference concerns the higher order discretization
in both time and space considered in \cite{Bryan2002}.
The second one 
is given by the numerical decoupling 
of acoustic waves considered in \cite{Bryan2002}. 
Figure \ref{fig:dry_128} (right) 
shows the same results for $\theta'$
and a $256\times 128$ grid, compared with a solution
computed using the same numerical scheme this time
on a finer grid of $512\times 256$ (consequently,
time steps are roughly halved to $0.1\,$s).
It can be thus seen that a higher resolution in both time and space
compensates for the lower order discretizations
and yields better agreement with the solution 
in \cite{Bryan2002}, as seen in Figure \ref{fig:dry_256}.
In particular,
maximum and minimum values of vertical velocity are this time
equal to 
$13.9372\,$m s$^{-1}$ and $-8.30085\,$m s$^{-1}$,
respectively.

\subsubsection{The Moist Simulation}\label{Subsec:Moist}

Here we consider the same configuration as above, 
but now with a moist atmospheric environment.
A neutrally stable environment can be obtained by considering 
the wet equivalent potential temperature $\theta_e$,
defined for a reversible moist adiabatic atmosphere by
\begin{equation}\label{eqn:theta_eq}
\theta_e = T \left( 
\frac{p_a}{p_{00}}
\right)^{-R_a/(\cpa +\cpl r_t)}
\exp
\left[
\frac{L_v r_v}{(\cpa + \cpl r_t) T}
\right],
\end{equation}
taken from \cite{Emanuel94}.
Supposing that the total water mixing ratio
is constant at all levels,
the vertical profiles of $\pi$, $\theta$, $r_v$,
and $r_c$ can be obtained using 
(\ref{eqn:pi0}), (\ref{eqn:thetarho}),
and (\ref{eqn:theta_eq}),
if values for $\theta_e$ and $r_t$
are provided.
We finally compute the hydrostatic
base state written in terms of $p$, $T$, $q_v$, and $q_l$
in our formulation.
The value of $r_t$ must be greater than
$\rvs=\qvstar/q_a$, so that
the initial environment is saturated,
that is, $q_v = \qvstar$ and $q_l >0$
everywhere in the domain.
The initial perturbation (\ref{eqn:dry_pert})
is then introduced in such a way that the buoyancy fields
are identical in both the dry and moist simulations, 
when $\theta_0 = 300\,$K
in the dry case \cite{Bryan2002}.
The initial field for $\theta$ is thus given by
\begin{equation*}\label{eqn:theta_ini}
\theta (p,T) \left(1 +
\frac{\rvs(p,T)}{\epsilon}
\right) =
\theta_{\rho0} (1+r_t) \left(
\frac{\theta'}{300} + 1
\right),
\end{equation*}
which is solved pointwise for $T$
throughout the domain at constant pressure $p(z)$.
\begin{figure}[!ht]
\centering
\includegraphics[width=0.49\textwidth]{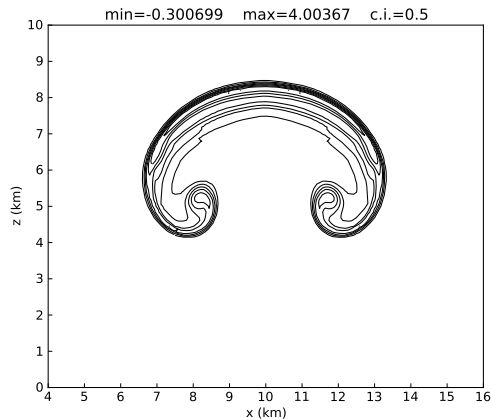}
\caption{Moist thermal simulation at $1000\,$s.
Perturbation potential temperature on a $256\times 128$ grid,
contoured every $0.5\,$K.
}
\label{fig:moist_128}
\end{figure}
\begin{figure}[!ht]
\noindent
\includegraphics[width=0.49\textwidth]{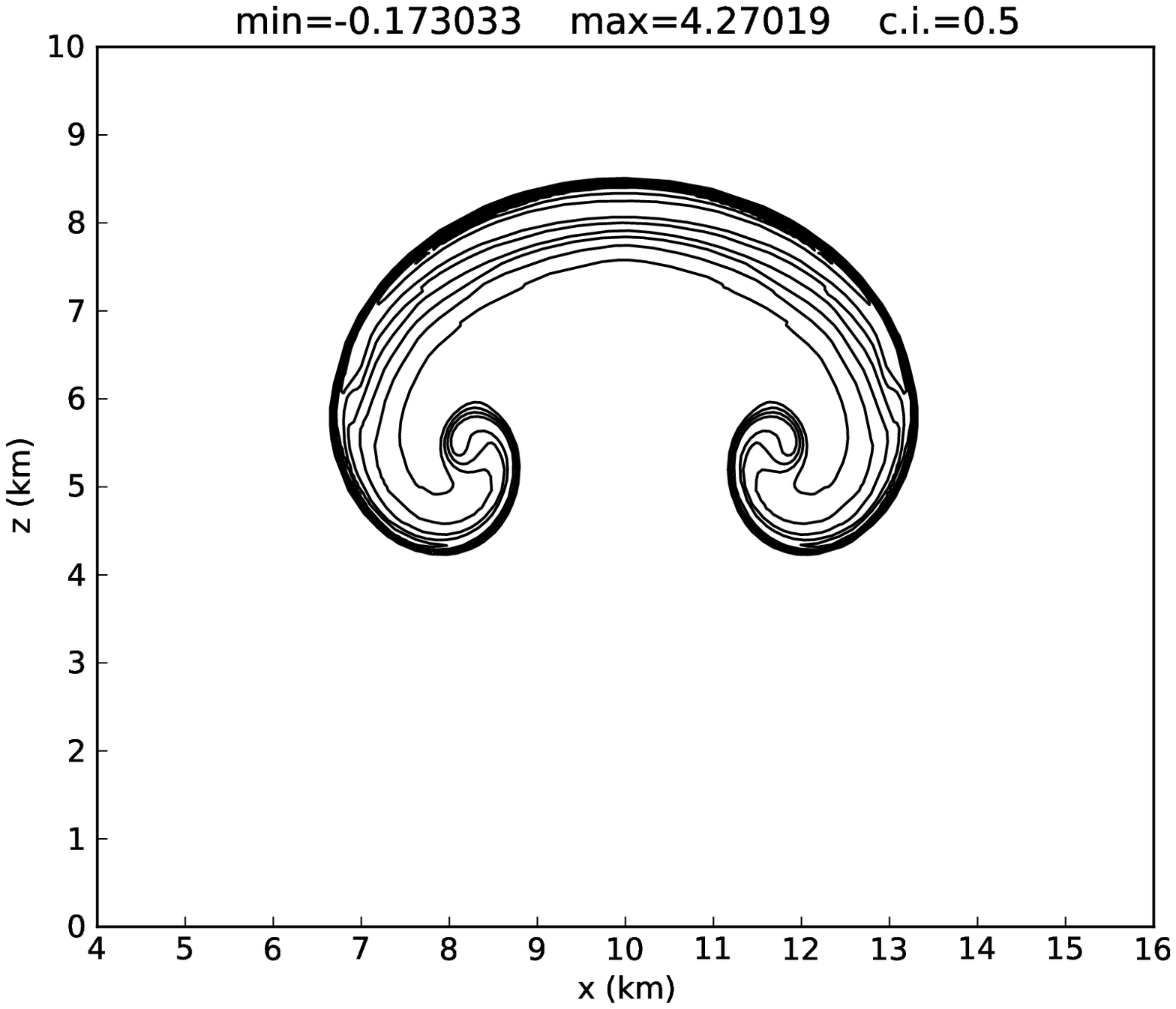}
\includegraphics[width=0.49\textwidth]{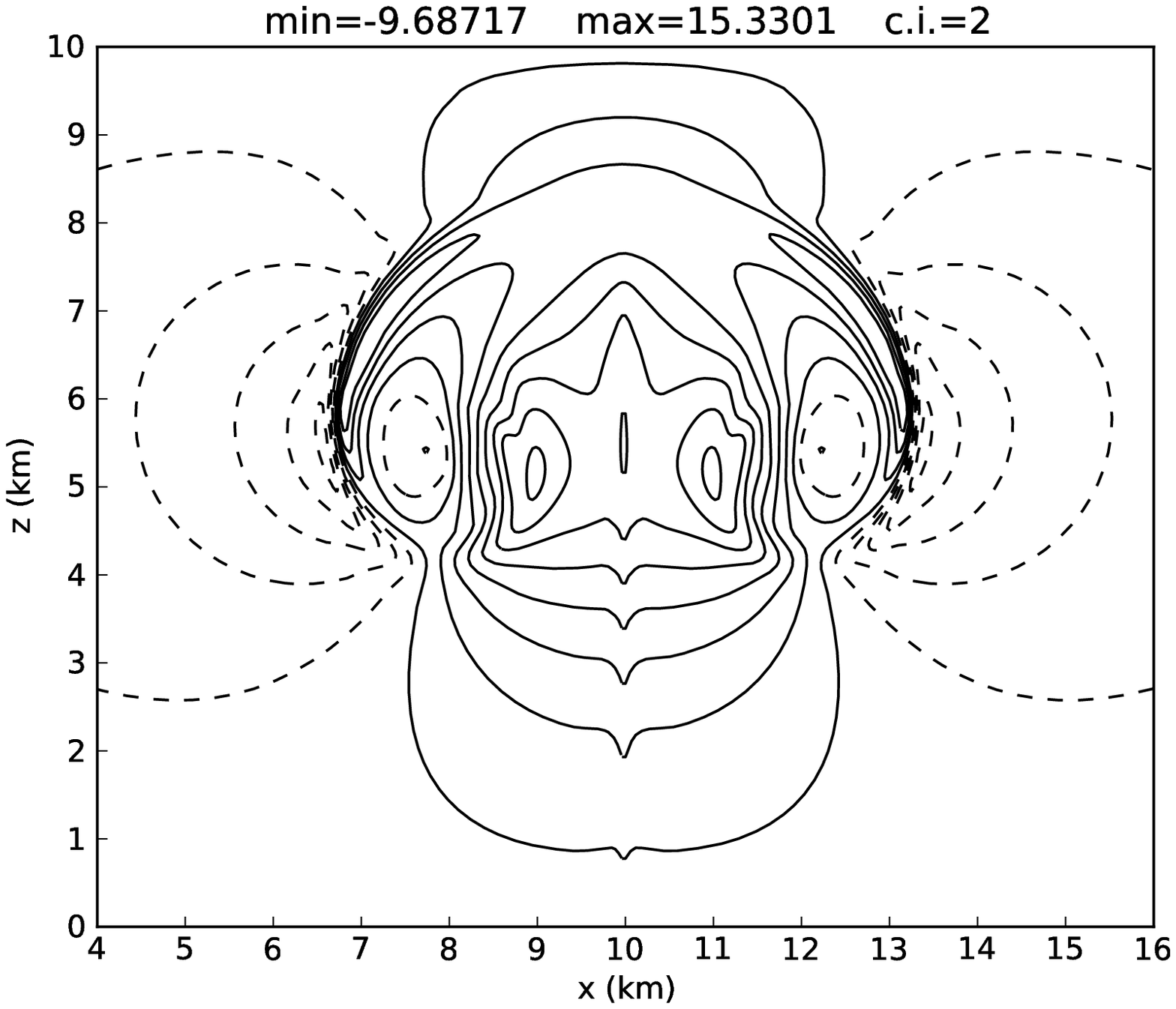}\\
\caption{Moist thermal simulation at $1000\,$s on a $512\times 256$ grid.
Left: perturbation potential temperature
contoured every $0.5\,$K.
Right: vertical velocity contoured every $2\,$m s$^{-1}$,
negative contours are dashed.
Contrast to Fig.~3 in \cite{Bryan2002}.}
\label{fig:moist_256}
\end{figure}

First, we use the one-step coupled scheme described in 
\S~\ref{Subsec:Coupled} 
to compute the moist rising thermal with input
parameters: $\theta_{e0}=320\,$K and $r_t=0.02$.
For a $256\times 128$ grid,
time steps are 
roughly constant of about $0.21\,$s,
similar to the dry computation.
The introduction of moist microphysics involves
an additional cost of approximately $15$ to $20\,$\% in CPU time
with respect to the dry computation,
which roughly corresponds to the computational
cost of the Newton iterative procedure, 
described through {\bf Steps 1-4} 
in Section \ref{Sec:Num_Met},
to solve the nonlinear system (\ref{eqn:solveT})
throughout the domain.  (A fixed tolerance $tol = 10^{-10}$ 
has been used for the Newton solver for all results shown.)
The maximum and minimum values for 
the perturbation wet equivalent potential temperature
($\theta'_e = \theta_e - \theta_{e0}$) 
are 
given by $4.00367\,$K and $-0.300699\,$K, respectively,
compared with the original 
$4.09521\,$K and $-0.305695\,$K in \cite{Bryan2002}.
Our computation yields 
$13.3267\,$m s$^{-1}$ and $-8.77365\,$m s$^{-1}$,
for the maximum and minimum vertical velocities, respectively,
which are slightly lower than  
$15.7130\,$m s$^{-1}$ and $-9.92698\,$m s$^{-1}$
in \cite{Bryan2002}.
Both solutions look reasonably similar in terms
of position, height, and width of the thermal,
as seen in Figure \ref{fig:moist_128}. 
Some differences can nevertheless be observed 
around the vortex cores, as in the previous
dry computation.  Increasing the spatial and temporal resolution 
as before yields even better agreement, as seen in 
Figure \ref{fig:moist_256} for a $512\times 256$ grid.
For instance, the maximum and minimum values of vertical velocity 
are now $15.3301\,$m s$^{-1}$ and $-9.68717\,$m s$^{-1}$,
respectively.
These results provide a validation of the numerical implementation of 
the dynamics solver in conjunction with the moist thermodynamics.

\subsection{Comparison of Different Schemes}

We now investigate the performance
of the two-step schemes detailed in 
\S~\ref{Subsec:Split} for the moist thermal simulation.
In both two-step schemes, the semi-split and the fully-split,
the dynamics is advanced allowing no conversion between water vapor
and liquid water for a time interval of 
$\dtsat$ before an adjustment step is performed to account for the 
saturation requirements.
For the purposes of this study, we now consider
the reference solution to be that given by the
one-step coupled scheme.  We focus on
evaluating the impact of $\dtsat$ on the 
results from the split schemes.

\subsubsection{Benchmark Problem}
We consider again the moist configuration
of the benchmark problem in \cite{Bryan2002}.
All simulations were carried out on a uniform grid of $256\times 128$.
Recall that the motivation for this study arises from the fact
that numerical methods that do not explicitly resolve the acoustic
modes typically run with a much larger time step than that required
by explicit evolution of the fully compressible equations.  Although
we continue to resolve the acoustic waves explicitly in this study,
we mimic the effect of these larger time steps on the representation of 
phase changes by increasing $\dtsat.$
\begin{figure}[!ht]
\noindent
\includegraphics[width=0.49\textwidth]{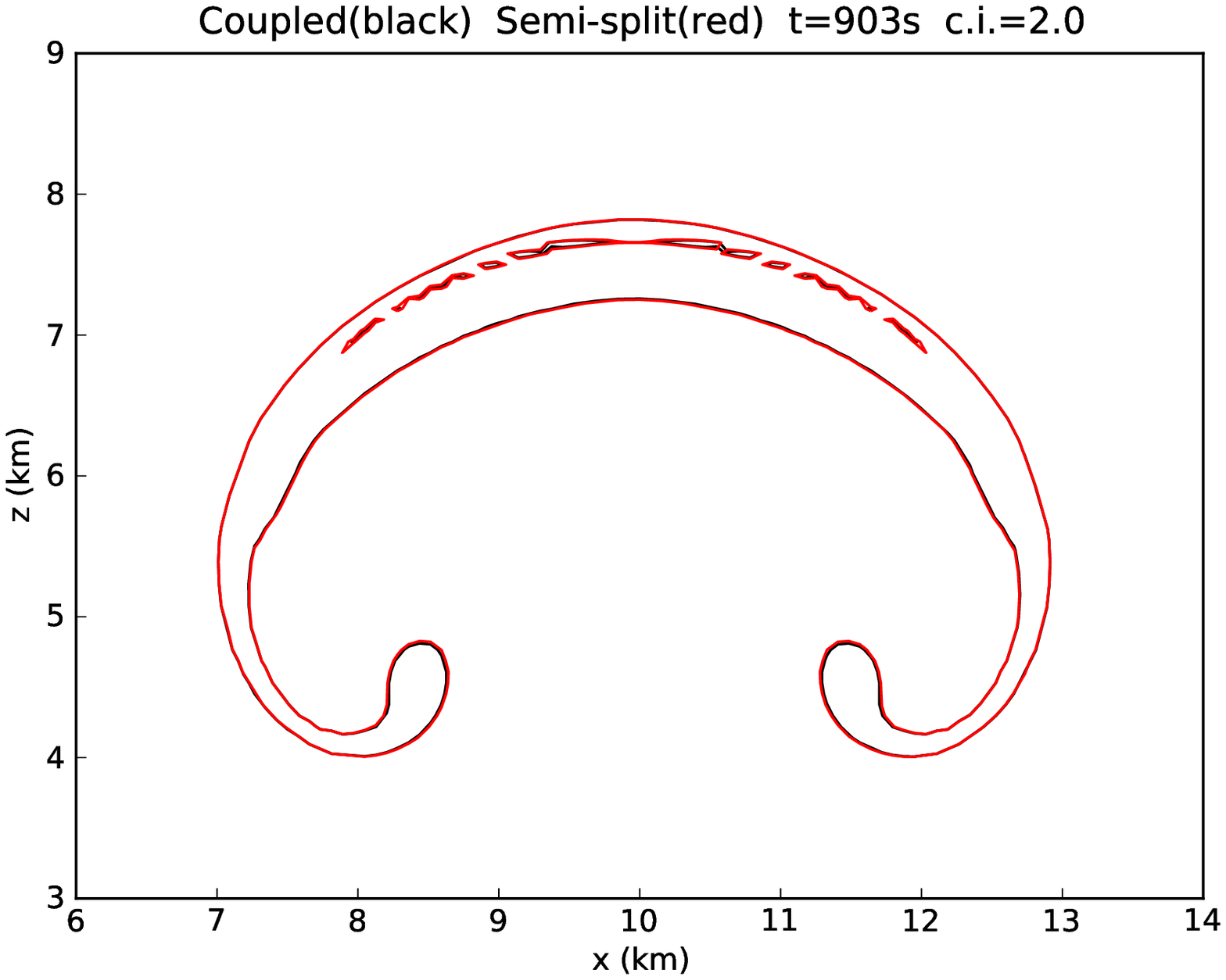}
\includegraphics[width=0.49\textwidth]{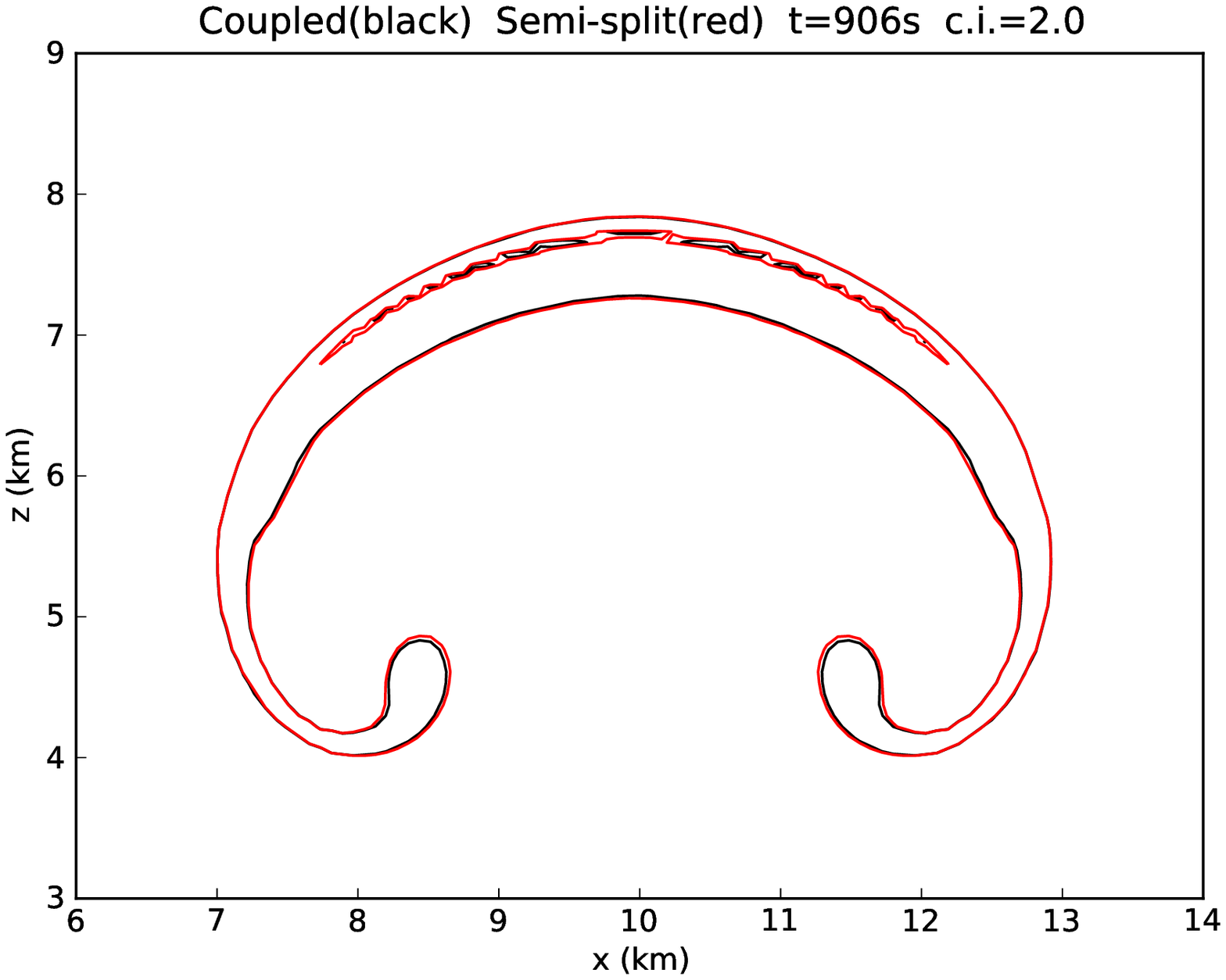}\\
\includegraphics[width=0.49\textwidth]{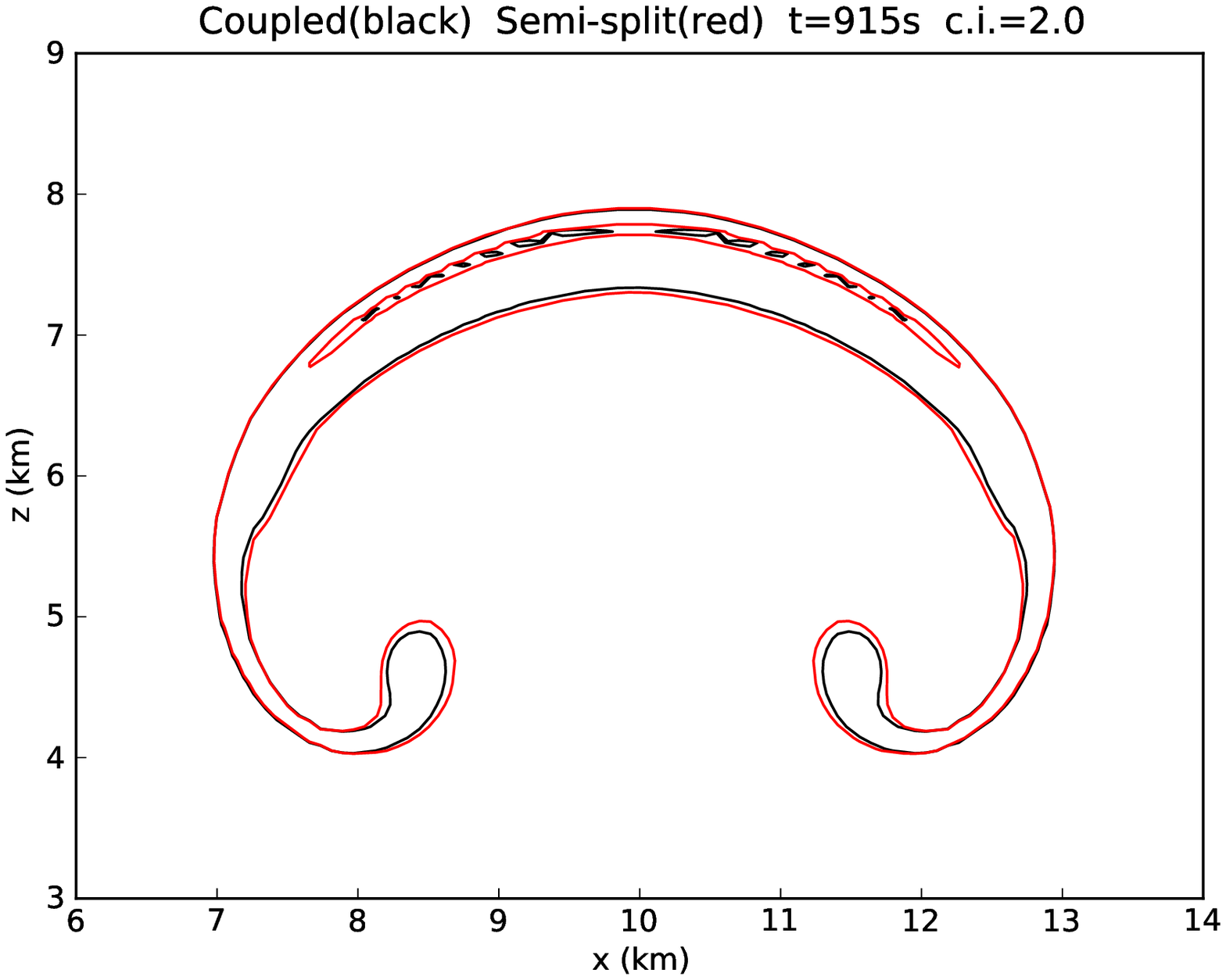}
\includegraphics[width=0.49\textwidth]{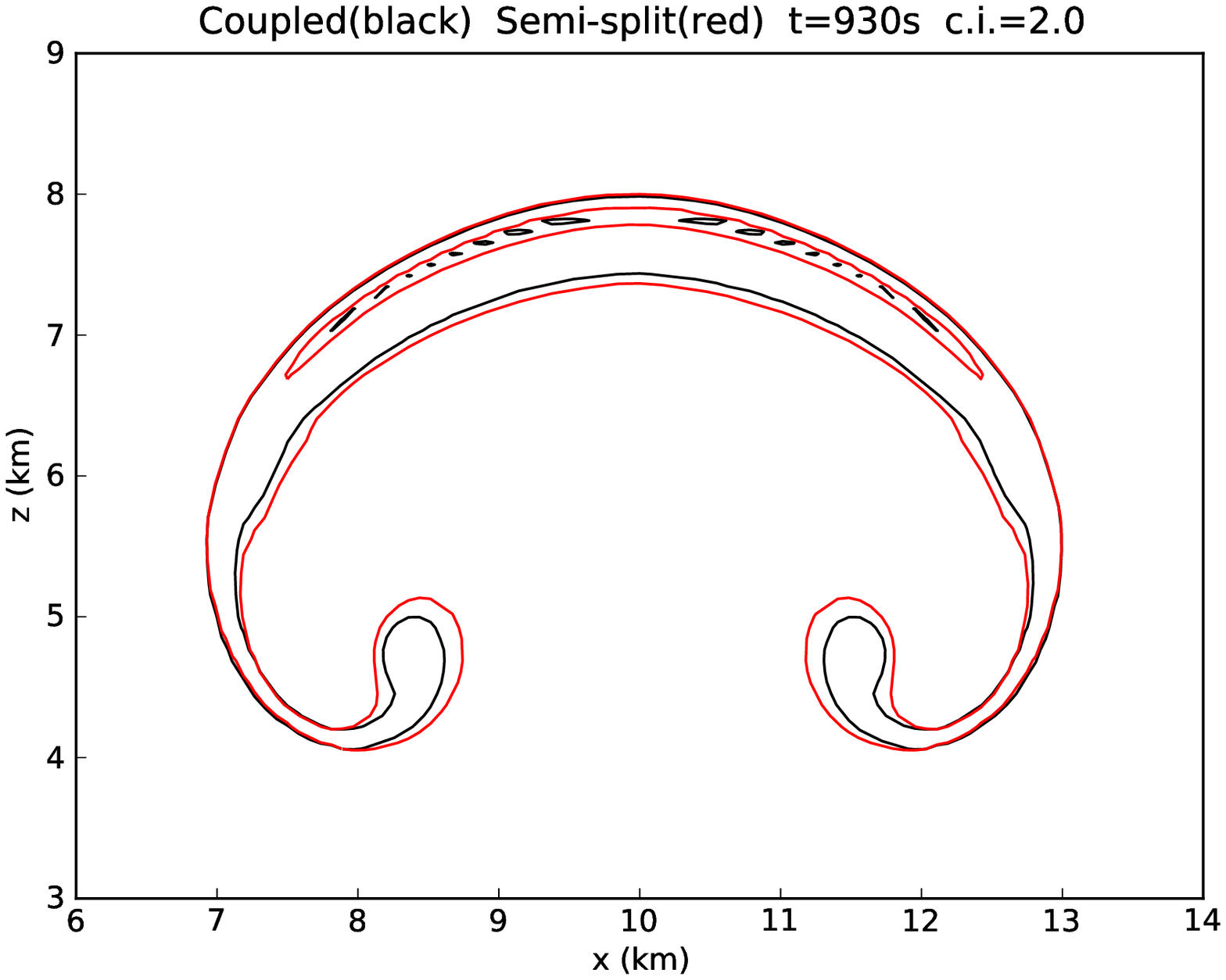}\\
\caption{
Semi-split solutions for $\theta'_e$
with $\dtsat=30\,$s and $\tsat=900\,$s
each overlaid on the reference solution. 
Top: $t=903\,$s (left) and $t=906\,$s (right).
Bottom: $t=915\,$s (left) and $t=930\,$s (right).
Contours every $2\,$K.}
\label{fig:comp_moist_semisplit}
\end{figure}
\begin{figure}[!ht]
\centering
\includegraphics[width=0.49\textwidth]{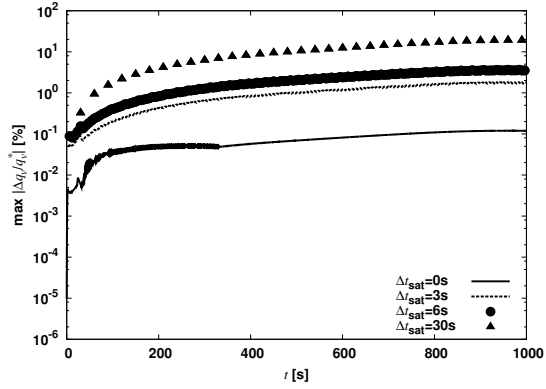}
\caption{
Drift of $q_v$ computed with the semi-split solver.
Time variation of $|\Delta q_v/\qvstar|$ in percentage.}
\label{fig:semisplit_drift}
\end{figure}
\begin{figure}[!ht]
\noindent
\includegraphics[width=0.49\textwidth]{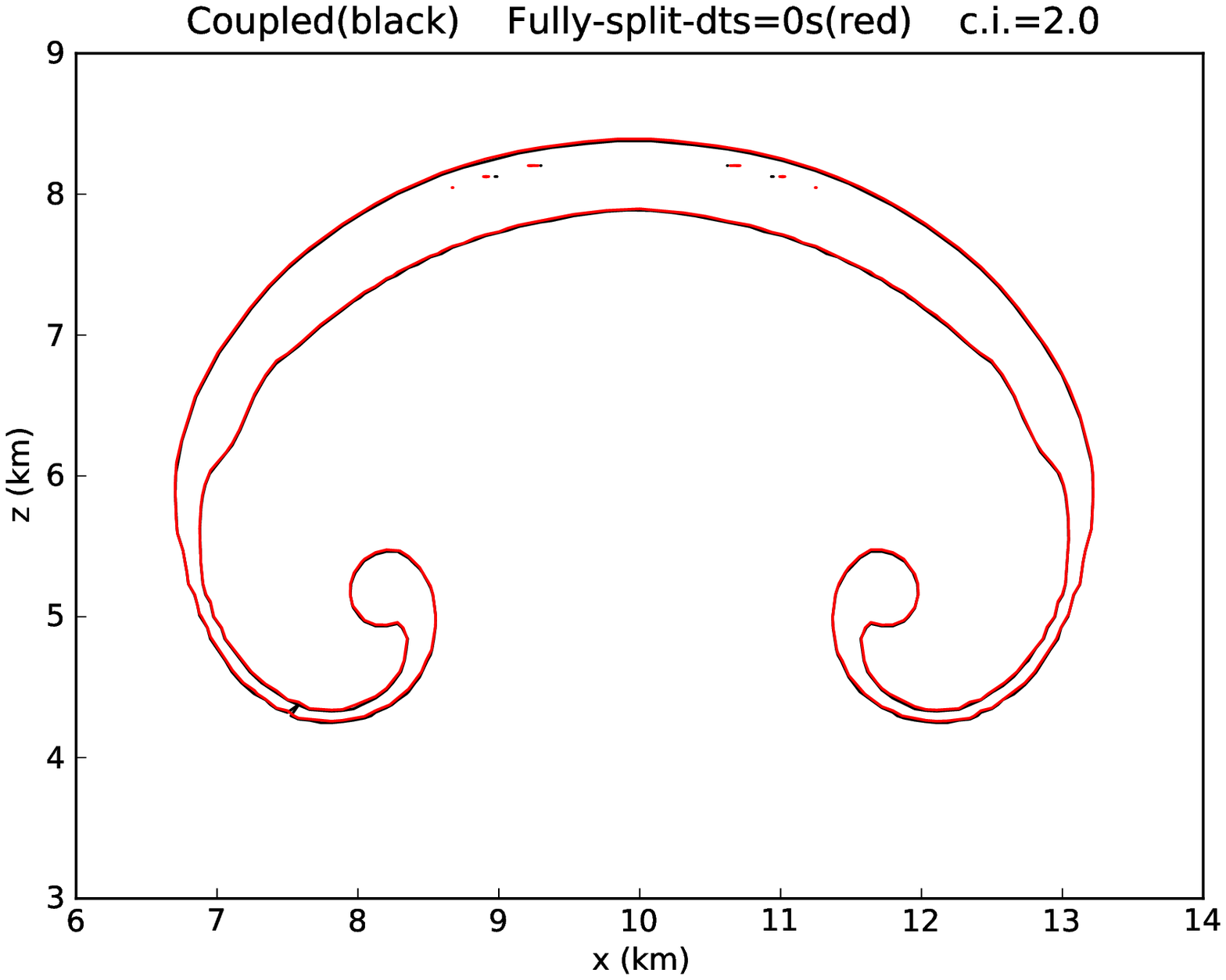}
\includegraphics[width=0.49\textwidth]{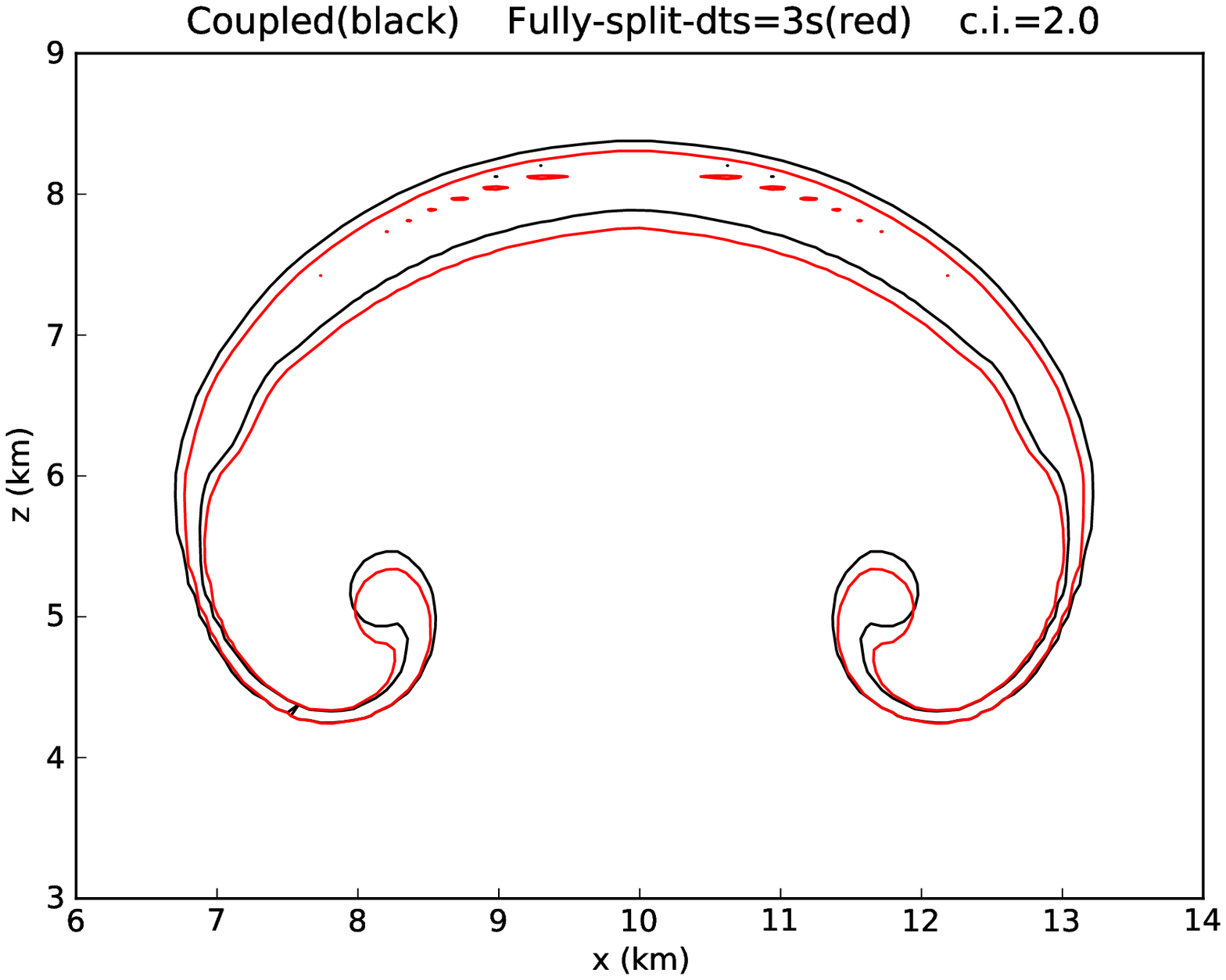}\\
\includegraphics[width=0.49\textwidth]{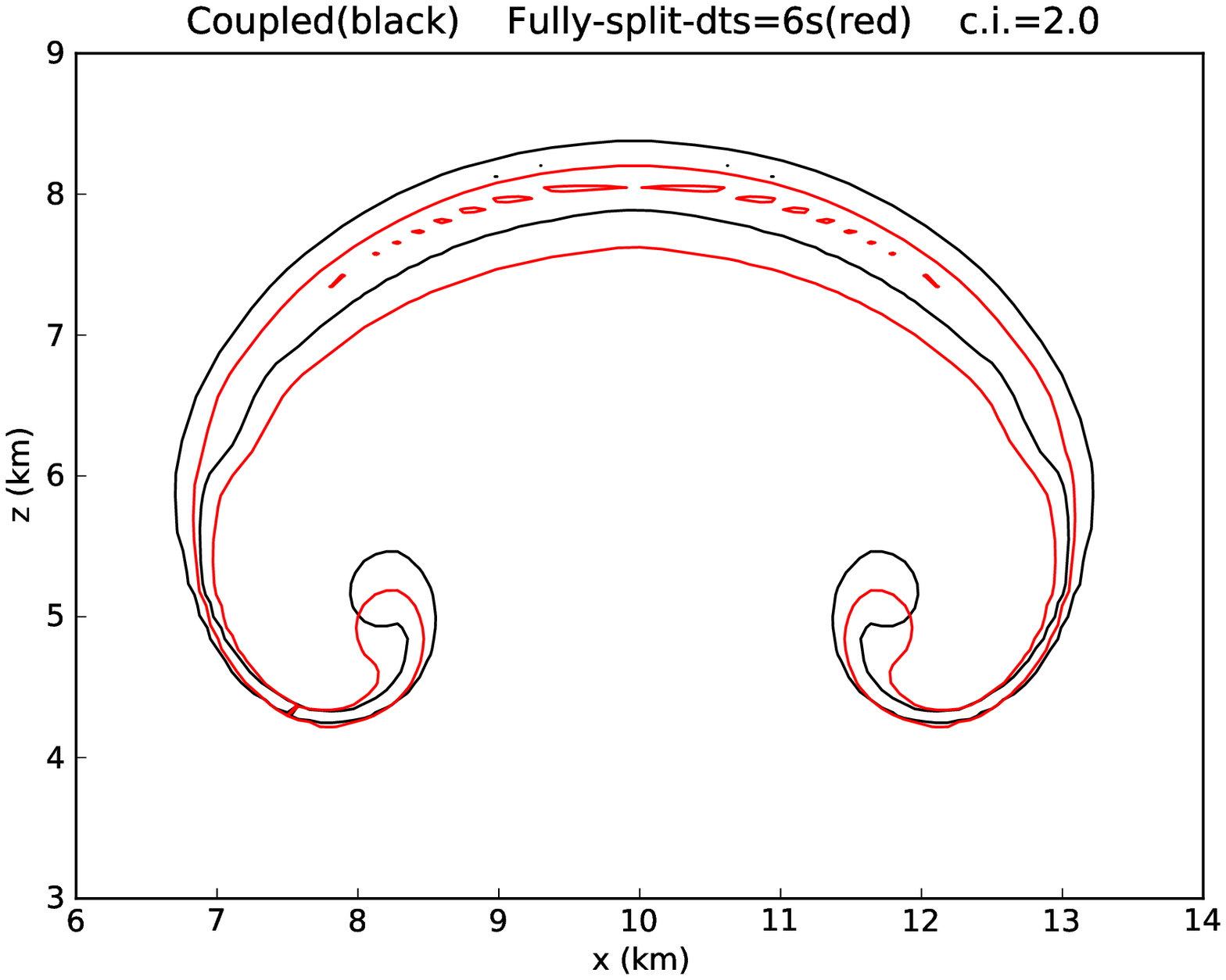}
\includegraphics[width=0.49\textwidth]{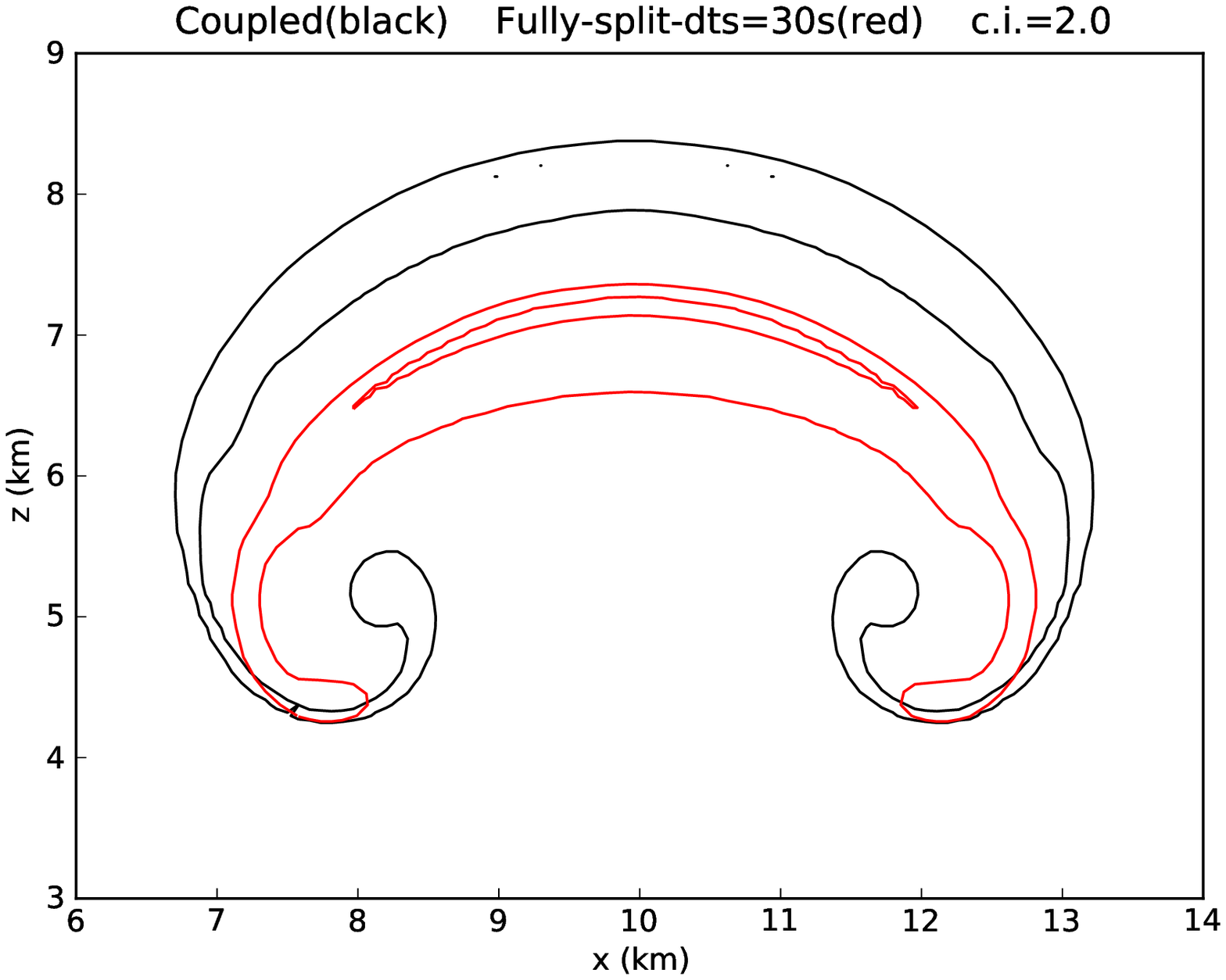}\\
\caption{
Fully-split solutions for $\theta'_e$
compared with the coupled ones.
Top: $\dtsat=0$ (left) and $\dtsat=3\,$s (right).
Bottom: $\dtsat=6\,$s (left) and $\dtsat=30\,$s (right).
Contours every $2\,$K.}
\label{fig:comp_moist_fullysplit}
\end{figure}

We first consider the effect of varying $\dtsat$ in the 
semi-split solver.  Regardless of the value of $\dtsat,$
the compressible dynamics are still evolved
with time steps of about $0.21\,$s.
For comparison, we note the time step 
used by \cite{ONeill2013} to solve the same problem 
in a pseudo-incompressible framework was $\dt = 1.66\,$s 
(corresponding to an advective CFL of $0.5$). 
As previously noted, the evolution of $\rho$, $\vel$, and $E$ 
is identical to that in the one-step coupled scheme.
The evolution of liquid water and water vapor neglects phase change,
therefore a drift in the values of $q_v$ and $q_l$ is observed with respect 
to the reference solution in which the saturation requirements are
verified
every time step.  
The semi-split solution drifts from the 
reference solution, 
corresponding in this case to the saturated state
$\qvstar$
and $q_w-\qvstar$, 
only until $t>\tsat+\dtsat$, at which point $q_v$ and $q_l$
are restored to the same values as in the reference solution,
since the dynamics of the semi-split solution are unaffected by the
drift.
In Figure \ref{fig:comp_moist_semisplit}
we present results for $\theta'_e$ 
at times $t=903$, $906$, $915$, and $930\,$s,
from a simulation with $\dtsat = 30\,$s and $\tsat = 900\,$s
each overlaid on the reference solution.
For time $930\,$s, Figure \ref{fig:comp_moist_semisplit}
shows the semi-split solution before the saturation adjustment has
taken place.
Not surprisingly, the larger the time since the last saturation 
adjustment at $900\,$s, the larger the differences 
between the semi-split solution and the reference solution.
After the adjustment step both solutions are identical.
Recall that 
even though the dynamics 
and thus the position of the thermals
are the same in both cases,
$\theta_e$ depends on $q_v$ and $q_l$, 
via $r_v$ and $T$ in (\ref{eqn:theta_eq}),
which are not the same in both solutions 
between times $900$ and $930\,$s.

Defining $\Delta q_v = q_v - \qvstar$ as the maximum value of the
drift over each $\dtsat$ (which occurs when we reach 
$t^{n+1} \geq \tsat + \dtsat$),  we show in Figure \ref{fig:semisplit_drift}
the variation of $|\Delta q_v/\qvstar|$ in percentage
for simulations with $\dtsat = 0$, $3$, $6$, and $30\,$s.
Again, not surprisingly, we observe that the maximum drift 
is roughly proportional to $\dtsat$; 
when $\dtsat = 3\,$s the maximum drift is almost 
$2\,$\%, whereas for $\dtsat = 30\,$s the drift reaches almost
$20\,$\%.  For this particular problem, the maximum drifts 
are due to a local excess of the computed $q_v$ with
respect to its saturated value.
In the simplified set of equations considered here, 
$q_v$ and $q_l$ are not used in any other microphysical processes,
thus there is no practical impact from the error due to the drift.
However, in a more realistic simulation in which the values of $q_v$ and $q_l$
might enter into other processes, the results here demonstrate that 
if $\dtsat \gg \dt,$ one must be cautious in the use of $q_v$ and $q_l$ 
with lagged saturation adjustment, even if the dynamics is correctly described.

We now consider the effect of $\dtsat \gg \dt$ on the 
evolution of the dynamics using the fully-split solver.
In Figure \ref{fig:comp_moist_fullysplit} 
we present results from simulations using the fully-split solver
and $\dtsat = 0$, $3$, $6$, and $30\,$s,
again each overlaid on the reference solution. 
Here we observe that fully neglecting the effect of phase change
on the dynamics for long time intervals (relative to the 
time step numerically defined by the acoustics) 
allows significant deviations from the reference
solution.
To quantify this difference, we note that the maximum vertical velocities 
obtained with the fully-split solver are $13.3267$, $12.9247$, $12.6430$, 
and $9.40192\,$m s$^{-1}$, for $\dtsat = 0$, $3$, $6$, and $30\,$s, 
respectively; whereas the semi-split solver yields 
$13.3267\,$m s$^{-1}$ in all cases, consistent with the coupled reference solution.

\subsubsection{Non-isentropic Background State}

We consider the hydrostatically balanced profiles
in \cite{Clark1984} (Eq.~2) for the background state:
\begin{equation}\label{eq:back_pb2}
\left.
\begin{array}{l}
\theta_0 (z) = \theta_{00} \exp(Sz), \\[1.5ex]
\ds
p_0(z) = p_{00} \left[ 1 - \frac{g}{\cpa \theta_{00} S} 
\left(
1 - \exp(-Sz)
\right)
\right]^{\cpa/R_a},
\end{array}
\right\}
\end{equation}
where $\theta_{00}$ and $p_{00}$
stand for the environmental potential temperature
and pressure at the surface ($z=0$),
with the static stability $S$ defined as
$S= N^2/g = d \ln \theta_0/dz$
($N$ is the Brunt-V\"ais\"al\"a frequency).
The potential temperature is given by (\ref{eqn:pi_theta}).
For the following computations
we define a computational domain $4\,$km
high and wide, with periodic horizontal 
boundary conditions and the same vertical 
boundary conditions implemented for the previous benchmark
problem.
The same thermodynamic parameters from \cite{Bryan2002}
are considered, whereas 
constants (\ref{eqn:constant_pvstar}) 
coming from \cite{Romps2008}
are considered in the
Clausius--Clapeyron equation (\ref{eqn:pvstar})
with $\ptrip = 611\,$Pa.
From \cite{GrabowskiClark1991}, 
we take $S=1.3\times 10^{-5}\,$m$^{-1}$,
$\theta_{00}=283\,$K, and $p_{00}=850\,$hPa.
All simulations were performed on a uniform
grid of $256\times 256$.

For our formulation
we also need to compute the hydrostatic base density
$\rho_0$ based on the background temperature
and pressure (\ref{eq:back_pb2}),
and the distribution of air,
water vapor, and liquid water in the atmosphere.
The latter quantities are set by the
relative humidity in the atmosphere RH
measured in percentage
and defined as RH$~= (p_v/\pvstar)\times 100$.
In particular if RH$_0 <100\,$\%,
then no liquid water should be present in the 
atmosphere in order to guarantee the thermodynamic 
equilibrium of the initial state, that is, $q_{l0}(z)=0$.
We consider in this study two cases:
first, a saturated medium, that is
RH$_0 =100\,$\% and $r_t = 0.02$, just like in the
moist benchmark problem;
and a second configuration with RH$_0 =20\,$\%,
and hence, no liquid water in the initial 
background state.
Contrary to the benchmark configuration in 
\cite{Bryan2002}, we now have in either case
a non-isentropic background state, where
the following definitions
of specific entropy have been adopted
\cite{Romps2008}:
\begin{eqnarray}
 s_a & = & \cpa \log \left( 
\frac{T}{\Ttrip}
\right) -
R_a \log \left( 
\frac{p}{\ptrip}
\right),
\nonumber
\\[1.5ex]
s_v & = & \cpv \log \left( 
\frac{T}{\Ttrip}
\right) -
R_v \log \left( 
\frac{p}{\ptrip}
\right) + \Sov,
\nonumber
\\[1.5ex]
s_l & = & \cvl \log \left( 
\frac{T}{\Ttrip}
\right),
\nonumber
\\[1.5ex]
s_m & = & q_a s_a + q_v s_v + q_l s_l, \label{eq:entropy_m}
\end{eqnarray}
for dry air, water vapor, liquid water,
and moist air, with 
$\Sov = \Eov/\Ttrip + R_v$.
(Notice that the
specific entropy of dry air at the triple point is 
neglected in this definition.)
\begin{figure}[!ht]
\noindent
\includegraphics[width=0.49\textwidth]{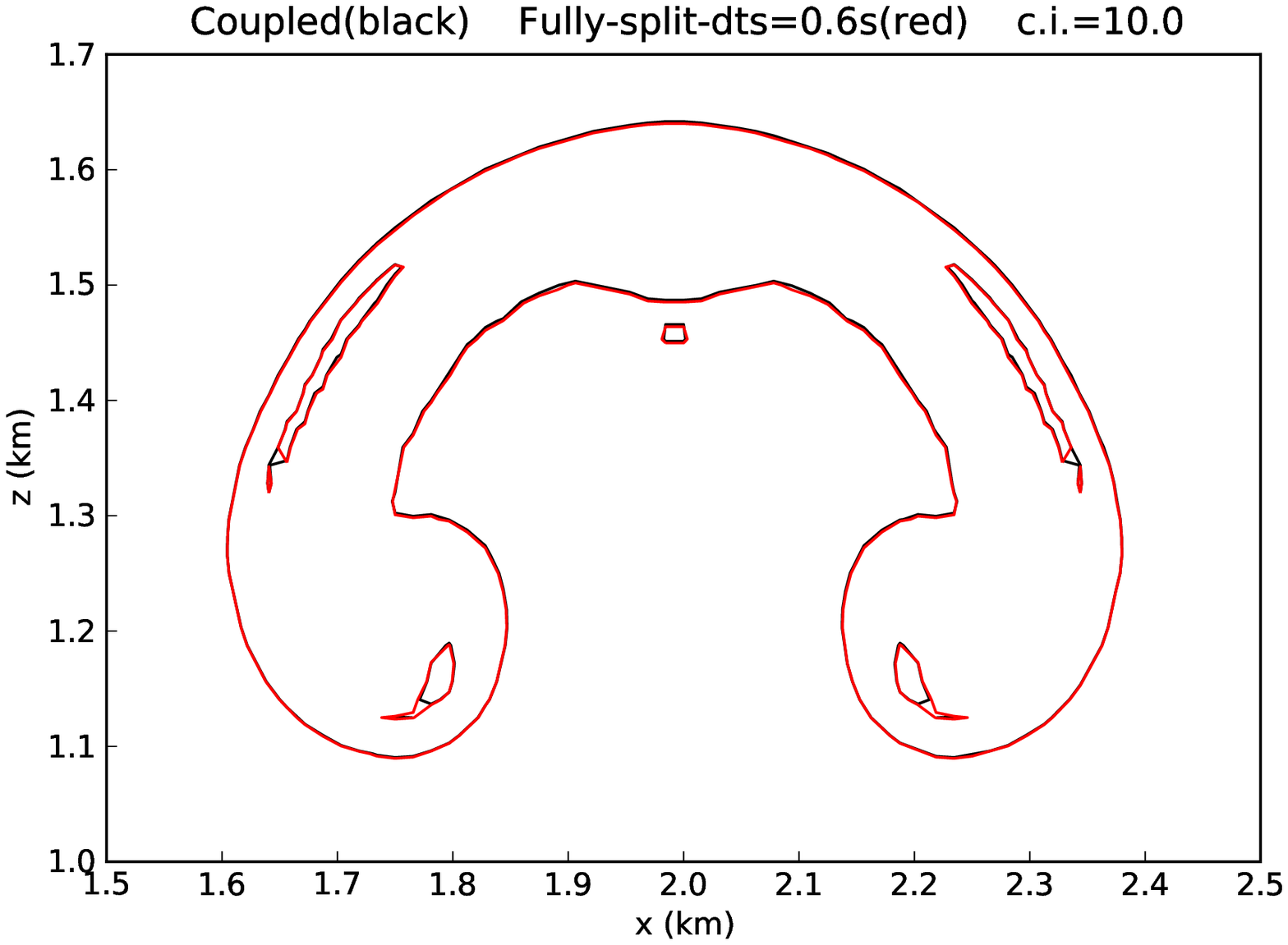}
\includegraphics[width=0.49\textwidth]{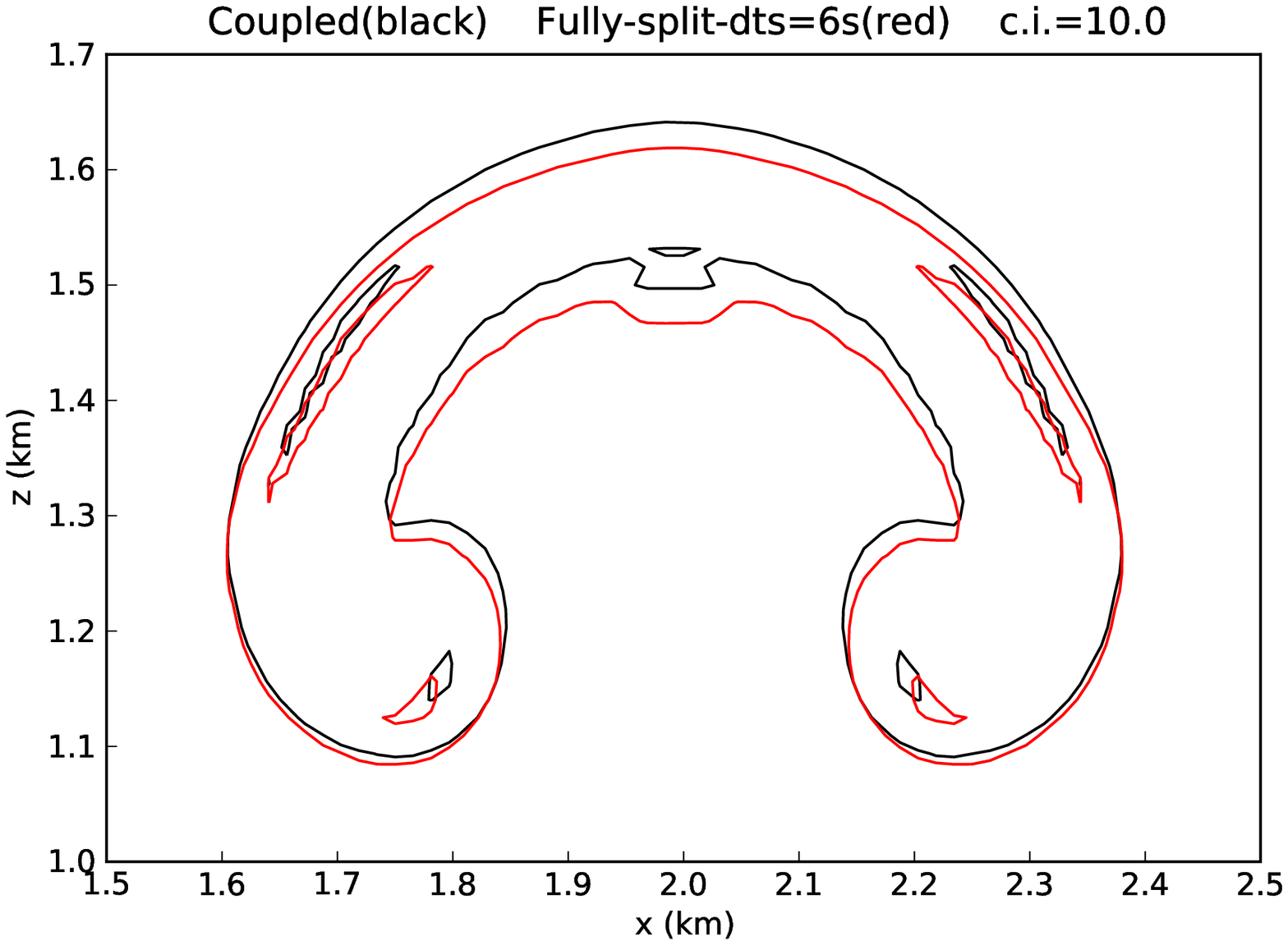}\\
\caption{Initially saturated, non-isentropic background state.
Fully-split 
solutions for $s_m$
after $300\,$s
with $\dtsat=0.6\,$s (left) and $\dtsat=6\,$s (right),
compared with the coupled ones.
Contours every $10\,$J kg$^{-1}$ K$^{-1}$.}
\label{fig:comp_moist_pb2}
\end{figure}

Let us consider the first configuration 
with an initially saturated environment.
In this case the moist 
squared Brunt-V\"ais\"al\"a frequency
$N_m^2$ 
can be computed according to
\cite{Durran1982} (eqs. 36--37),
which yields 
$N_m^2$ 
monotonically varying
from 
$3.5\times 10^{-6}\,$s$^{-2}$
at the surface
up to
$1.3\times 10^{-5}\,$s$^{-2}$
at the top of the computational domain.
Positive values of $N_m^2$ imply static moist stability.
Similar to (\ref{eqn:dry_pert}),
we introduce a warm perturbation on temperature:
\begin{equation}\label{eqn:temp_pert}
T' = 2 \cos ^2 \left( \frac{\pi L}{2} \right),
\end{equation}
where $L$ is defined by (\ref{eq:pert_L}),
with 
$x_c = 2\,$km,
$z_c = 0.8\,$km, and
$x_r = z_r = 300\,$m.
The water distributions, as well as the density,
are thus adjusted to
the perturbed temperature with 
the original pressure field.
As for the previous problem
both the coupled
and the split schemes
with $\dtsat = 0$,
yield 
the same 
solutions,
with $\sigma^\mathrm{CFL}=0.9$
and roughly constant time steps of about $0.04\,$s.
Increasing $\dtsat$ in the fully-split scheme 
has a comparable effect to that seen
in the benchmark problem with an isentropic base state.
Figure \ref{fig:comp_moist_pb2} illustrates
the latter behavior in terms of the specific
entropy of moist air (eq. (\ref{eq:entropy_m}))
for  $\dtsat = 0.6$ and $6\,$s,
after $300\,$s of integration.
Simulations were stopped
before the nonlinearities become more apparent 
and sub-grid turbulence starts playing a more
important role in the dynamics, as
analyzed in \cite{GrabowskiClark1991}.
We recall that for the sake of simplicity
sub-grid turbulence is not considered in the present study.
We note that the choice of $\dtsat = 0.6\,$s is based on the approximate 
size of the time step that would be used if computed from the advective
rather than acoustic CFL condition, i.e. if the time step were based
on the fluid velocity rather than the sound speed.

\begin{figure}[!ht]
\centering
\includegraphics[width=0.49\textwidth]{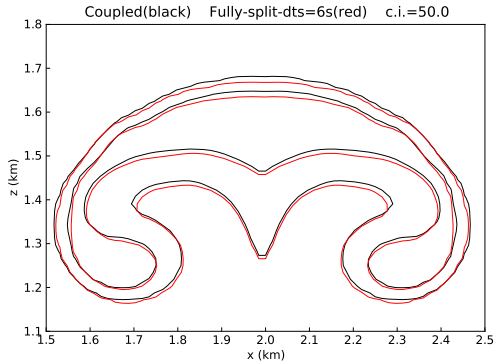}\\
\caption{Non-isentropic background state with a partially
saturated perturbation.
Fully-split 
solution for $s_m$
after $300\,$s
with $\dtsat=6\,$s 
compared with the coupled one.
Contours every $50\,$J kg$^{-1}$ K$^{-1}$.}
\label{fig:comp_moist_pb3}
\end{figure}

For the second configuration with RH$_0 =20\,$\%,
we consider the same temperature perturbation
(\ref{eqn:temp_pert})
and an additional circular perturbation 
on the relative humidity, which is 
set to $100\,$\% for a radius $r < 200\,$m,
as considered in \cite{GrabowskiClark1991}.
A transition layer is assumed such that 
\begin{equation}\label{eqn:rh_pert}
 {\rm RH} = {\rm RH}_0 + (100 - {\rm RH}_0)
\cos ^2 \left( \frac{\pi }{2}
\frac{r-200}{100}
 \right),
\
200 \leq r \leq 300,
\end{equation}
taken also from \cite{GrabowskiClark1991}.
Initially there is no liquid water in the domain,
not even in the saturated region, 
whereas the perturbed water vapor is recomputed
based on (\ref{eqn:temp_pert}) and
(\ref{eqn:rh_pert}) with the original static
pressure.
After performing the same tests
with the different numerical techniques,
the same observations can be made 
in terms of moist 
saturation
adjustments.
For instance, Figure \ref{fig:comp_moist_pb3}
shows results 
obtained with the fully-split scheme
with an adjustment interval of 
$\dtsat = 6\,$s;
once again $\sigma^\mathrm{CFL}=0.9$,
which yields roughly constant time steps of about $0.04\,$s.
Notice that in this case, differences
between the fully-split and coupled approximations
are smaller with respect to the previous
configurations
for saturated and both isentropic and non-isentropic
environments.
This is due to the fact
that all the liquid water
is mainly contained in the perturbed area
and hence only this region is subjected to 
phase changes and active moist microphysics.

\section{Summary}

In this paper we have studied the incorporation of reversible
moist processes related to phase change phenomena into 
numerical simulations of atmospheric flows. 
Specifically, we have tried to characterize the impact of 
modifying the time scale at which the moist thermodynamics is
adjusted to the saturation requirements.
For the purpose of this study, the compressible Euler equations 
were written in a form including conservation equations for 
total density, momentum, and energy of moist air,  
and were explicitly evolved with time steps dictated by the 
acoustic CFL condition.

Two different approaches were considered to evolve the system.
In the first approach, a one-step coupled procedure solves 
the equations of motion
together with a conservation equation for total water content.
Because of the choice of variables,
in particular because the energy of moist air 
includes the contribution of both sensible and latent 
heats, this formulation does not include  source terms related 
to phase change in either the energy or the total water equation.
Therefore, the system of equations can be solved without needing 
to estimate or neglect source terms related to phase change.
The pressure used to update the momentum and energy in the evolution
equations is computed from the equation of state following a 
saturation adjustment procedure.

In the second approach, the evolution equation for total water
is replaced by separate evolution equations for liquid water and
water vapor, where source terms related to phase change now appear. 
A two-step technique is implemented in which the system of equations 
is first evolved with these source terms set to zero. 
In a second step, a saturation adjustment procedure is performed
after a time interval $\dtsat$, updating the values of liquid
water and water vapor.  We consider two variants of the two-step scheme.
In the first, a semi-split strategy in which the dynamics 
of the moist flow are correctly computed, a drift is expected 
and observed in the values of water vapor and liquid water during
the time interval in which the saturation adjustment is not imposed.
In the second, fully-split scheme, 
the saturation adjustment is not performed during the 
evolution of the dynamics,  and the dynamics themselves are
seem to drift from those of the fully coupled solution.

In summary, numerical tests of the semi-split scheme
showed that 
non-trivial deviations of the water vapor and 
liquid water from their correct values may occur 
even when the dynamics is correctly described. Tests of the
fully-split scheme demonstrated that imposing the saturation
adjustment too infrequently relative to the time step at which
the dynamics evolve
may lead to inaccuracies in the dynamical evolution.
Further testing with both isentropic and non-isentropic background states, 
as well as saturated and non-saturated initial configurations,
confirmed the initial findings.  
It is hoped that the insight gained here
as to how closely the saturation adjustment should be 
numerically coupled to the dynamics will carry over to methods
in which the dynamics themselves are evolved with larger time steps.
This will be further investigated and discussed
in future work.

\section*{Acknowledgments} 
The work in the Center for Computational Sciences and Engineering
at LBNL was supported by the Applied Mathematics Program of
the DOE Office of Advance Scientific Computing Research under
U.S. Department of Energy under contract No.~DE-AC02-05CH11231.
DR was supported by the Scientific Discovery through Advanced
Computing (SciDAC) program funded by U.S. Department of Energy 
Office of Advanced Scientific Computing Research and 
Office of Biological and Environmental Research.

\bibliographystyle{plain}
\bibliography{ws}

\clearpage

\end{document}